\documentclass[aps,pra,amsmath,twocolumn,amssymb,superscriptaddress]{revtex4}

\usepackage[dvips]{graphicx}
\usepackage{dcolumn}
\usepackage{bm}
\usepackage{dsfont}
\usepackage{color}

\usepackage{url}
\usepackage[colorlinks=true ,citecolor=blue,urlcolor=blue]{hyperref}
\usepackage{amsmath,amssymb}

\definecolor{Nathanblue}{rgb}{0.,0.24,0.51}
\newcommand{\blue}{\color{Nathanblue}}

\def\bs#1{\boldsymbol{#1}}

\renewcommand{\thefigure}{\arabic{figure}}

\begin{document}


\title{{\Large {\blue Direct imaging of topological edge states in cold-atom systems}}}

\author{N. Goldman}
\email{ngoldman@ulb.ac.be}
\affiliation{Center for Nonlinear Phenomena and Complex Systems - Universit\'e Libre de Bruxelles (U.L.B.), B-1050 Brussels, Belgium}
\author{J. Dalibard}
\affiliation{Laboratoire Kastler Brossel, CNRS, ENS, UPMC, 24 rue Lhomond, 75005 Paris}
\affiliation{Coll\`ege de France, 11, place Marcelin Berthelot, 75005 Paris, France}
\author{A. Dauphin}
\affiliation{Center for Nonlinear Phenomena and Complex Systems - Universit\'e Libre de Bruxelles (U.L.B.), B-1050 Brussels, Belgium}
\affiliation{Departamento de F\'isica Te\'orica I, Universidad Complutense, 28040 Madrid, Spain}
\author{F. Gerbier}
\affiliation{Laboratoire Kastler Brossel, CNRS, ENS, UPMC, 24 rue Lhomond, 75005 Paris}
\author{M. Lewenstein}
\affiliation{ICFO -- Institut de Ci\`encies Fot\`oniques, Parc Mediterrani de la Tecnologia, 08860 Barcelona, Spain}
\affiliation{ICREA -- Instituci\'o Catalana de Recerca i Estudis Avan\c cats, 08010 Barcelona, Spain}
\author{P. Zoller}
\affiliation{Institute for Quantum Optics and Quantum Information of the Austrian Academy of Sciences, A-6020 Innsbruck, Austria} \affiliation{Institute for Theoretical Physics, Innsbruck University, A-6020 Innsbruck, Austria}
\author{I. B. Spielman}
\affiliation{Joint Quantum Institute, National Institute of Standards and Technology, and University of Maryland, Gaithersburg, Maryland, 20899, USA}



\begin{abstract}  
Detecting topological order in cold-atom experiments is an ongoing challenge, the resolution of which offers novel perspectives on topological matter. In material systems, unambiguous signatures of topological order exist for topological insulators and quantum Hall devices. In quantum Hall systems, the quantized conductivity and the associated robust propagating edge modes -- guaranteed by the existence of non-trivial topological invariants -- have been observed through transport and spectroscopy measurements. Here, we show that optical-lattice-based experiments can be tailored to directly visualize the propagation of topological edge modes. Our method is rooted in the unique capability for initially \emph{shaping} the atomic gas, and imaging its time-evolution after suddenly removing the shaping potentials. Our scheme, applicable to an assortment of atomic topological phases, provides a method for imaging the dynamics of topological edge modes, directly revealing their angular velocity and spin structure.
 \end{abstract}

\maketitle

{\Large T}he integer quantum Hall (QH) effect revolutionized our understanding of quantum matter, revealing the existence of exotic phases not described by the standard theory of phase transitions \cite{HasanKane2010,qi2011}. In this phenomenon, the Hall conductivity is quantized, $\sigma_H = (e^2/h) \, \nu$ -- where $e$ is the electron charge, $h$ is PlanckÕs constant and $\nu$ is an integer -- whenever the Fermi energy resides in an energy gap.   The integers $\nu$ are related to topological invariants -- Chern numbers -- that are associated with the bulk energy bands \cite{HasanKane2010,Thouless1982,Kohmoto1989}.  Their topological origin guarantees that the Chern numbers are constant as long as the bulk gaps remain open, explaining the signature plateaus in the Hall resistivity, present when external parameters, such as magnetic fields, are varied. Moreover, a holographic principle stipulates that a topologically-ordered \emph{bulk} gap, with topological invariant $\nu$, necessarily hosts $\nu$ propagating modes localized on the samples \emph{edge} \cite{Hatsugai1993}.  These topological edge states are chiral -- their motion has a well-defined orientation --  inhibiting scattering processes in the presence of disorder. 

In condensed matter physics, direct observations of edge states remain relatively rare. A first signature was obtained from magnetoplasmons created by pulsed voltages \cite{Ashoori1992}. Another evidence arises from edge transport in engineered Aharonov-Bohm interferometers with QH systems \cite{Ji2003,Zhang2009}. By contrast the ``routinely used" spectroscopic reconstruction of mid-gap states \cite{HasanKane2010} is consistent with the expected topological band structure, but does not prove their chiral nature.

Cold atoms trapped in optical lattices and subjected to synthetic gauge fields \cite{Aidelsburger2011,dalibard2011a} are an ideal platform for realizing topological insulating phases.  Making topology manifest in experiments, however, is a fundamental challenge.  In this context, transport experiments are conceivable \cite{Brantut:2012} but technically demanding. Existing proposals for measuring topological invariants \cite{alba2011a,Zhao2011,Goldman:2012njp,Price2012,umucalilar2008a} have experimental drawbacks and can only be applied to very specific configurations. Likewise, detecting topological edge states \cite{Scarola2007,Goldman2010a,StanescuEA2010,Goldman:2012prl,Kraus2012,Killi2012}, e.g., through light scattering methods~\cite{Pap2008}, require complicated manipulations to separate the small edge-state signal from the bulk background \cite{Goldman:2012prl,Goldman:2012prlbis}. \\

\begin{figure}
	\includegraphics[width=1.\columnwidth]{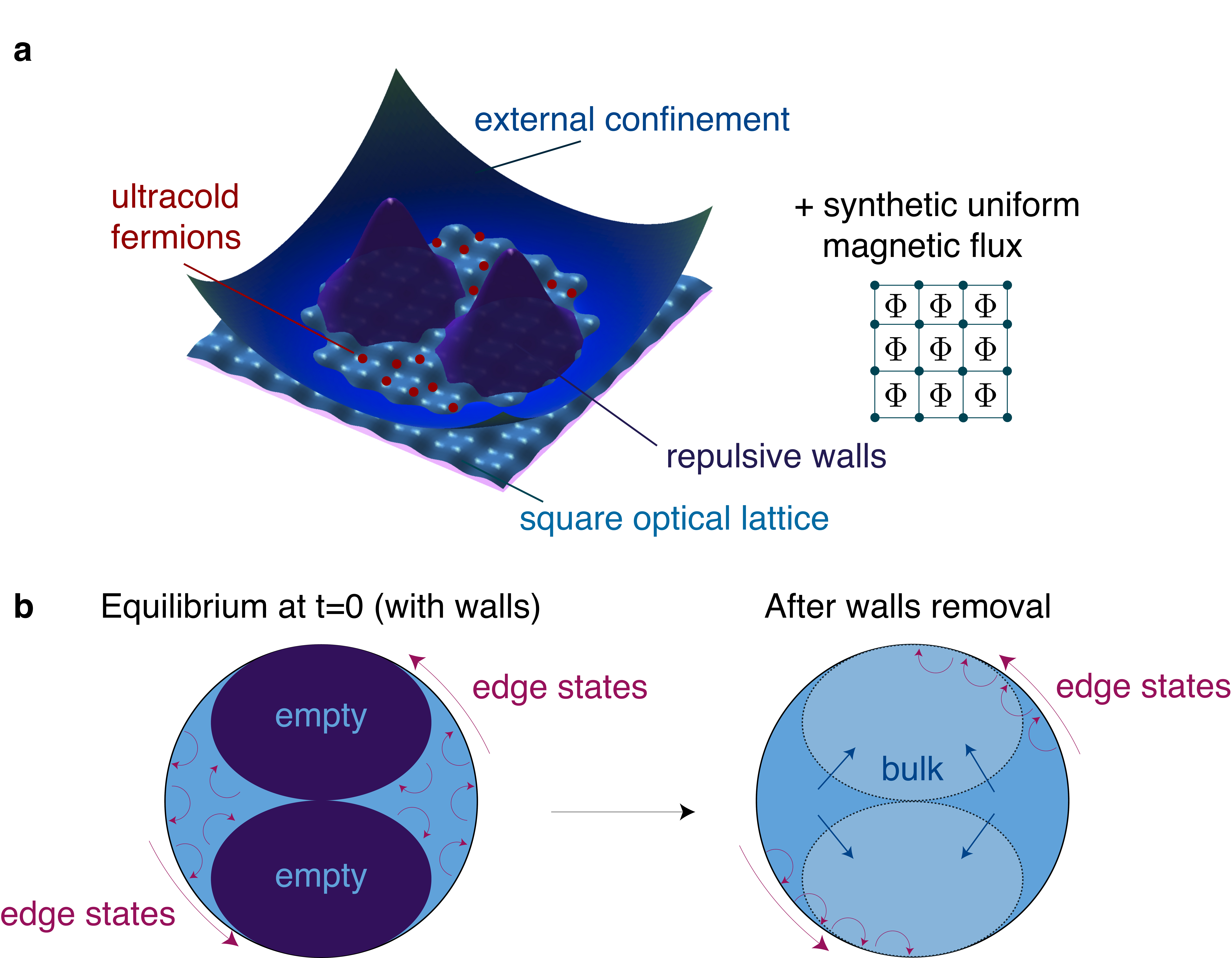}
	\caption{\label{figure1} {\bf Experimental scheme and general strategy.} {\bf a,}  Trapped cold atomic fermions move on a square optical lattice in the presence of a synthetic uniform magnetic flux $\Phi$.  Two repulsive potentials, initially forming holes in the atomic cloud, are suddenly removed at time $t=0$. At all times, atoms are confined by an additional circular potential. We generally assume that the confining barriers are perfectly sharp, but eventually discuss the case of smoother potentials. {\bf b,} The system is initially prepared in a quantum Hall phase: chiral edge states propagate along the edges determined by the repulsive walls and the external confinement. After releasing the walls, the edge states tend to propagate along the Fermi radius determined by the circular confinement: they encircle the initially vacant regions.}
\end{figure}

Here, we introduce a simple method to directly visualize the propagation of topological edge modes, by studying the time evolution of an atomic QH system after suddenly releasing constraining walls, see Fig.~\ref{figure1}. We show that the movement of the chiral edge states, encircling the initially vacant regions, is directly visible in the atomic density. This reveals the edge states' angular velocity and provides an unambiguous signature of chiral edge modes in the atomic system (Fig. \ref{figure1}). Our method is straightforward and insensitive to experimental imperfections: it only relies on a large initial occupancy of edge states. Crucially, our method requires that the edge states contribution to the density remains spatially separated from the bulk, which can be realized by populating a dispersionless bulk band with non-zero Chern number. We present several detection techniques, applicable to both flat and dispersive bands, that demonstrates the universal applicability to atomic systems with propagating edge modes. \\

\begin{figure}
 	\includegraphics[width=1\columnwidth]{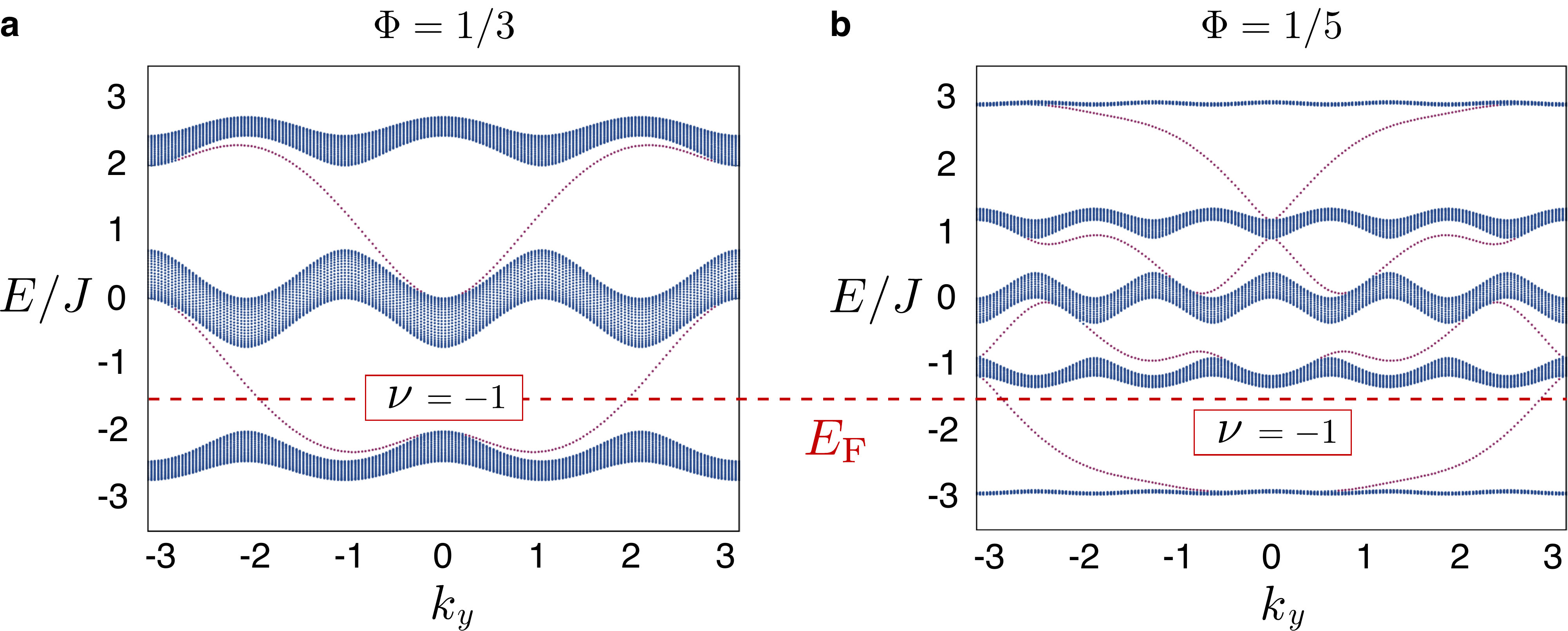}
	\caption{\label{figure2} {\bf Bulk and edge states spectrum: dispersive vs flat bands.} Energy spectrum $E(k_y)$ as a function of the quasi-momentum $k_y$ for {\bf a,}  $\Phi=1/3$ and {\bf b,} $\Phi=1/5$, obtained by diagonalizing the Hamiltonian \eqref{ham} on a finite cylinder directed along the $x$ direction, with $V_{\text{conf}}=0$. The projected bulk bands $E(k_x,k_y) \rightarrow E(k_y)$, shown in blue, are separated by large gaps of order $\sim J$. The red dispersion branches that are visible within the bulk gaps correspond to propagating modes that are localized on the opposite edges of the cylinder. When the Fermi energy is set within the first bulk gap, a single edge mode is populated on each edge of the cylinder (the lowest bulk band corresponds to the Chern number $\nu=-1$ for $\Phi=1/q$). When considering the circular geometry realized in an experiment ($V_{\text{conf}}\ne0$) and setting $E_{\text{F}}=-1.5J$, one is guaranteed that a single edge mode will be populated since the Chern number $\nu$ does not rely on the specific geometry used \cite{Goldman:2012prl,Goldman:2012prlbis,Buchhold2012}. In the case $\Phi=1/5$, the lowest energy band is characterized by the tiny flatness ratio, $f=W/\Delta \approx 0.04$, where $W (\Delta)$ denotes the first band (gap) width; in this topological quasi-flat band configuration, the populated edge states are expected to propagate more rapidly than the bulk.}
\end{figure}

We consider a two-dimensional optical lattice filled with non-interacting fermions, subjected to a uniform synthetic magnetic flux $\Phi$ \cite{jaksch2003a,gerbier2010a}, and confined by a circular potential, $V_{\text{conf}}(r)= V_0 (r/r_{0})^{\gamma}$.  In experiment, $V_{\text{conf}}(r)$ can be made nearly arbitrarily sharp $(\gamma\rightarrow\infty)$  \cite{Meyrath2005,Gaunt2012,Weitenberg2011}; this configuration is of particular interest for our scheme, as demonstrated below.  The resulting system realizes the Hofstadter model \cite{Hofstadter1976} with second-quantized Hamiltonian
\begin{align}
\hat H=& -J \sum_{m,n} \hat c^{\dagger}_{m+1,n} \hat c_{m,n} + e^{i 2 \pi \Phi m}  \hat c^{\dagger}_{m,n+1} \hat c_{m,n} + \text{h.c.} \notag \\
& + \sum_{m,n} V_{\text{conf}}(r)\, \hat c^{\dagger}_{m,n} \hat c_{m,n}.\label{ham}
\end{align}
$\hat c^\dagger_{m,n}$ describes the creation of a fermion at lattice site $\bs x /a=(m,n)$ where $m,n$ are integers; $J$ is the tunneling amplitude; and we take the lattice period $a$ as our unit of length.  This model has a topological band structure \cite{HasanKane2010,Kohmoto1989}: When $\Phi=p/q \in \mathbb{Q}$, the bulk energy spectrum splits into $q$ subbands \cite{Hofstadter1976}, each associated with a non-zero Chern number \cite{Kohmoto1989}. This guarantees the existence of robust edge states in the bulk energy gaps \cite{Hatsugai1993}. These edge states are \emph{chiral} in the sense that they propagate along the Fermi radius $R_{F}$ (i.e., the edge of the atomic cloud delimited by the confining potential $V_{\text{conf}}$) with a definite orientation of propagation. It is convenient to represent such non-trivial  spectra by diagonalizing the Hamiltonian \eqref{ham} on a cylindrical geometry  \cite{Hatsugai1993}, see Fig.~\ref{figure2}. This picture shows a clear separation of the bulk and edge states dispersions that survives in the experimental circular geometry produced by $V_{\text{conf}}$  \cite{Goldman:2012prl,Goldman:2012prlbis, Buchhold2012}. 

In the following, we specifically study the configurations $\Phi=1/3$ and $\Phi=1/5$ and set the Fermi energy $E_{\text{F}}=-1.5 J$ inside the lowest bulk energy gap, see Fig.~\ref{figure2}. In both these situations, the lowest energy band is associated with the Chern number $\nu=-1$, which guarantees the occupancy of a single edge mode with same chirality $\text{sign} (\nu)=(-)$. These two configurations differ significantly in that the occupied bulk band is \emph{nearly flat} (dispersionless) in the case $\Phi=1/5$, while it is dispersive for $\Phi=1/3$, see Fig. \ref{figure2}. \\

Our scheme \emph{(a)} demonstrates the existence of propagating modes that are localized close to the Fermi radius, and \emph{(b)} identifies their chirality and angular velocity $\dot{\theta}$.  To achieve this goal, we consider a geometry which constrains the QH system within two regions of the trap, as sketched in Figure \ref{figure1}, resembling a bat in flight.  This initial ``bat'' geometry is shaped by a pair of sharp potential walls $V_{\text{hole}}=V_{\text{hole 1}}+V_{\text{hole 2}}$ defined by $(x \pm r_0/2)^2 + (y/\sqrt{2})^2 <  r_0^2/4$, creating holes in the density distribution (see also Appendix \ref{app:size}). In the bat geometry, we set the Fermi energy within the lowest bulk gap $E_{\text{F}}=-1.5 J$ and suddenly remove $V_{\text{hole}}$ at time $t=0$. We then study the dynamics of the atomic density with all other parameters unchanged. The bat shape is optimized for visualizing the time-evolving chiral edge states in the density $\rho(\bs x, t)$ for $t>0$, see Fig. \ref{figure1}{\bf b}. In the following, we discuss how this ``wall-removal'' strategy can be exploited to reveal the edge states properties, as they progressively encircle the initially empty regions in a chiral manner.   

\section*{{\blue {\Large Results}}}

\section{Time-evolving density for dispersive systems}
 
\begin{figure}[h!]
 	\includegraphics[width=1\columnwidth]{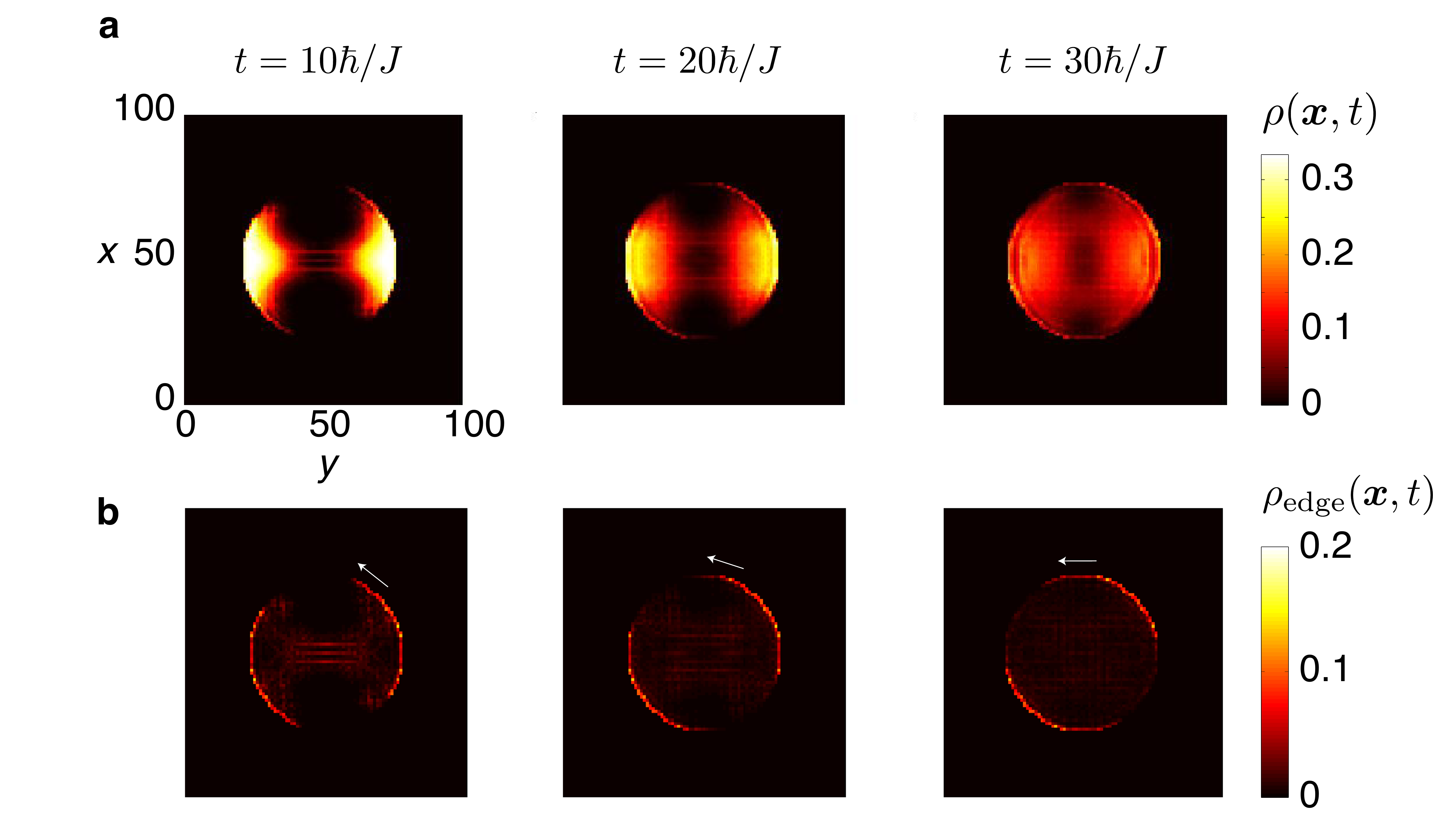}
	\caption{\label{figure3} {\bf Evolution of the spatial densities after releasing the walls.} {\bf a}, the spatial density $\rho (\bs x , t)$, and {\bf b}, the contribution of the initially populated edge states $\rho_{\text{edge}} (\bs x , t)$. The chiral motion is a signature of the non-trivial Chern number $\nu \ne 0$. In all the figures, $\Phi=1/3$, $E_{\text{F}}=-1.5 J$ and we considered infinitely sharp circular confinement ($r_0=27 a$) and ellipsoidal walls $V_{\text{hole}}$. The total number of particles is $N_{\text{part}}=210$, while the number of initially populated edge states is $N_{\text{edge}} \approx 80$. }
 \end{figure}

Figure \ref{figure3}{\bf a} shows the time-evolving density distribution $\rho (\bs x,t)$ for $\Phi=1/3$.  This example highlights the importance of the bulk band structure, as it demonstrates the drawbacks encountered when preparing a system with \emph{dispersive} bulk bands, see Fig. \ref{figure2}{\bf a}. The time evolution in Fig. \ref{figure3}{\bf a}, illustrates two main effects: (1) the progressive encircling of the holes by particles at the system's radial-edge (with localization length $\sim a$), and (2) the undesired and rapid filling of the holes by bulk states, see also Fig. \ref{figure1}{\bf b}. Once $t\approx10-20 \hbar/J$, the atomic cloud's initial bat shape has  become cyclonic, already indicating the presence of chiral edge states. 
To provide further insight, we separately calculated the contribution stemming from the initially populated edge states,  $\rho_{\text{edge}}(\bs x , t)$, see Appendix~\ref{app:one}. In the corresponding Fig. \ref{figure3}{\bf b}, we observe that the edge states, which propagated along the edges delimited by the bat potential at $t<0$, become localized along the circular edge at $r=R_F$, and that they follow a ``chiral'' motion. These edge states remain localized on the edge for very long times, and only slightly disperse into the bulk of the system, as can be anticipated from the small wavefunction overlap  between edge and bulk states.  

Figure \ref{figure3}{\bf a} emphasizes the problematic (non-chiral) filling of the holes by the many dispersive bulk states, which strongly inhibits the detection of the edge states in experiments.  The speed at which this filling occurs is to be compared with the circular motion of the edge states, which can be estimated from the group velocities $v \!=\! (1/\hbar) \partial E/\partial k$ associated with the bulk and edge states, see Fig. \ref{figure2}{\bf a}.  Additional interference takes place within the system, leading to small but visible patterns in the density. Figure \ref{figure3}{\bf b}, shows that these patterns mainly originate from interferences between bulk states. 
 
\begin{figure}[h!]
	\includegraphics[width=1\columnwidth]{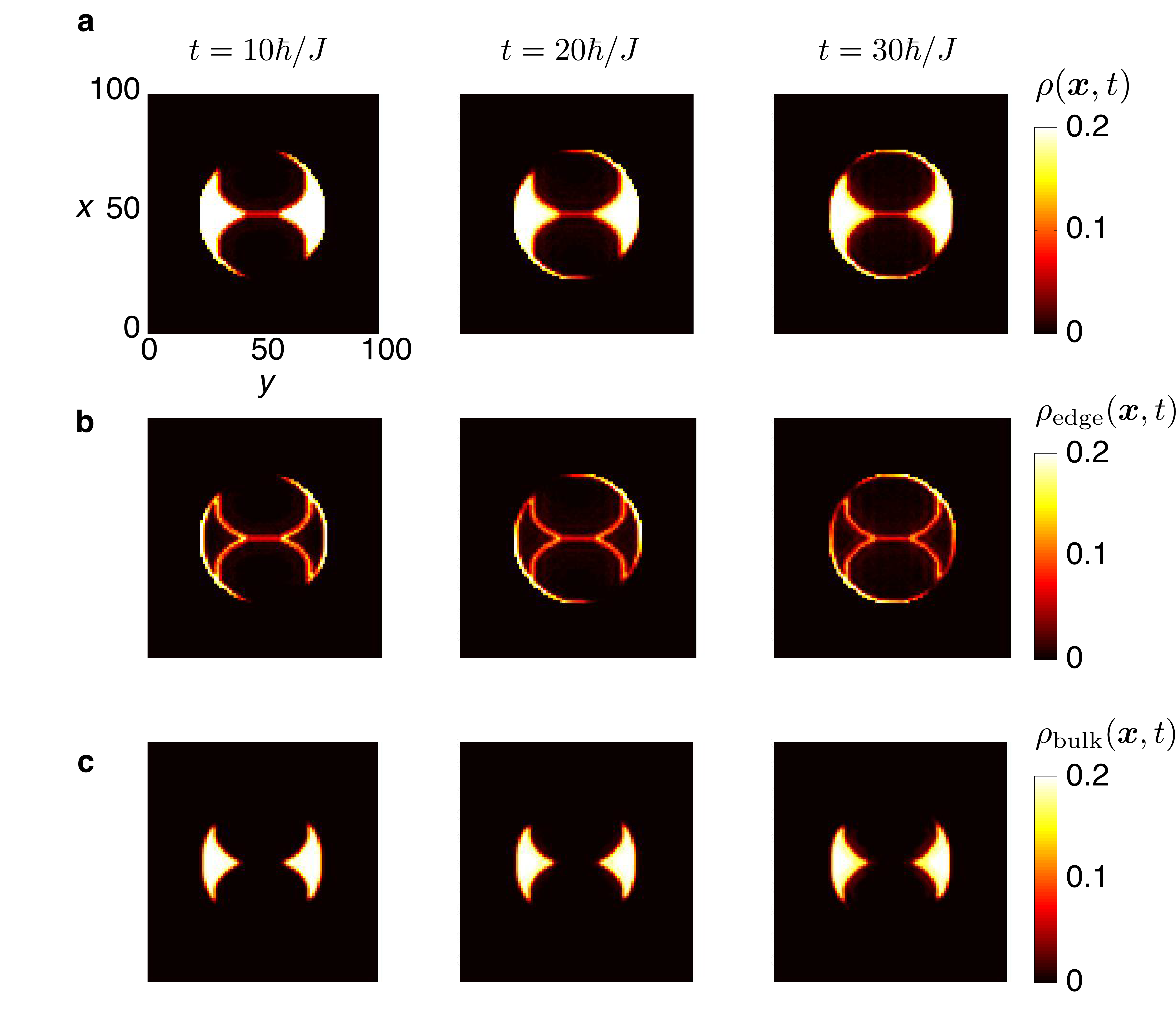}
	\caption{\label{figure4} {\bf The topological quasi-flat band configuration.} {\bf a}, the spatial density $\rho (\bs x , t)$, {\bf b}, the contribution of the initially populated edge states $\rho_{\text{edge}} (\bs x , t)$, and {\bf c}, the contribution of the initially populated bulk states $\rho_{\text{bulk}} (\bs x , t)$. In all the figures, $\Phi=1/5$ and $E_{\text{F}}=-1.5 J$. The total number of particles is $N_{\text{part}}=146$, while the number of initially populated edge states is $N_{\text{edge}} \approx 64$.  Note the \emph{dispersionless} nature of the occupied bulk states, which highly improves the detection of the edge state signal.
	}
\end{figure}
 
\section{Flat topological bulk bands}
For $\Phi=1/5$ and $E_{\text{F}}=-1.5 J$, the dispersionless (flat) bulk band represented in Fig. \ref{figure2}{\bf b} is totally filled, and most of the edge states lying in the first bulk gap are populated. The corresponding time-evolving density $\rho (\bs x,t)$, depicted in Fig. \ref{figure4}{\bf a}, is radically different than for $\Phi=1/3$, cf. Fig. \ref{figure3}{\bf a}. For $\Phi=1/5$, the edge states encircle the initially forbidden regions in a chiral manner, largely unperturbed by the now motionless bulk, making them directly visible in in-situ images of the cloud.  The dispersionless nature of the bulk states is further illustrated in Fig. \ref{figure4}{\bf c}, which shows the evolution of $\rho_{\text{bulk}}(\bs x , t)=\rho(\bs x ,t)-\rho_{\text{edge}}(\bs x , t)$. The initial bulk states are immobile for times $\sim 30 \hbar/J$. In Fig. \ref{figure4}{\bf b}, the evolution of the initially populated edge states $\rho_{\text{edge}}(\bs x , t)$ shows an interesting behavior: some edge states with energies close to the flat bulk band are almost dispersionless, and remain localized on the edges delimited by the bat potential. In contrast, the many edge states at higher energies are dispersive, encircling the holes in a clear and chiral manner. The instructive dynamics of the density $\rho (\bs x,t)$, which is due to the clear separation of the edge and bulk states during the evolution, can also be understood by studying the population of the single-particle eigenstates (see Appendix \ref{app:one}). Moreover, our method is highly robust against perturbations in the density (or equivalently in the Fermi energy, $E_{\text{F}} \approx -1.5 J + \delta$), as it only relies on the occupation of dispersionless bulk states and sufficiently many edge states.  We verified that the edge state signal is unambiguous when a high energy (dispersive) band is initially filled (see Appendix \ref{app:sensitivity}). 

Thanks to the topological quasi-flat band configuration, the edge states propagation can be visualized on long time scales, without being affected by the bulk dispersion. For $\Phi=1/5$ and a typical system size $R_F \sim 100 a$, we find $\dot \theta \sim 0.01 J / \hbar$, which would require a realistic time $t\approx 70 \hbar/J$ to observe the $\pi/4$-rotation undergone by the edge states. 

We verified that our results are stable when the spacious ellipsoidal walls are replaced by small perturbative potentials (see Appendix \ref{app:size}). Finally, the edge/bulk ratio can be further improved by  initially confining the entire atomic cloud to a small region located in the vicinity of the circular edge (see Appendix \ref{app:size}).

\section{Revealing topological edge states in dispersive systems}
\begin{figure}
	\includegraphics[width=1\columnwidth]{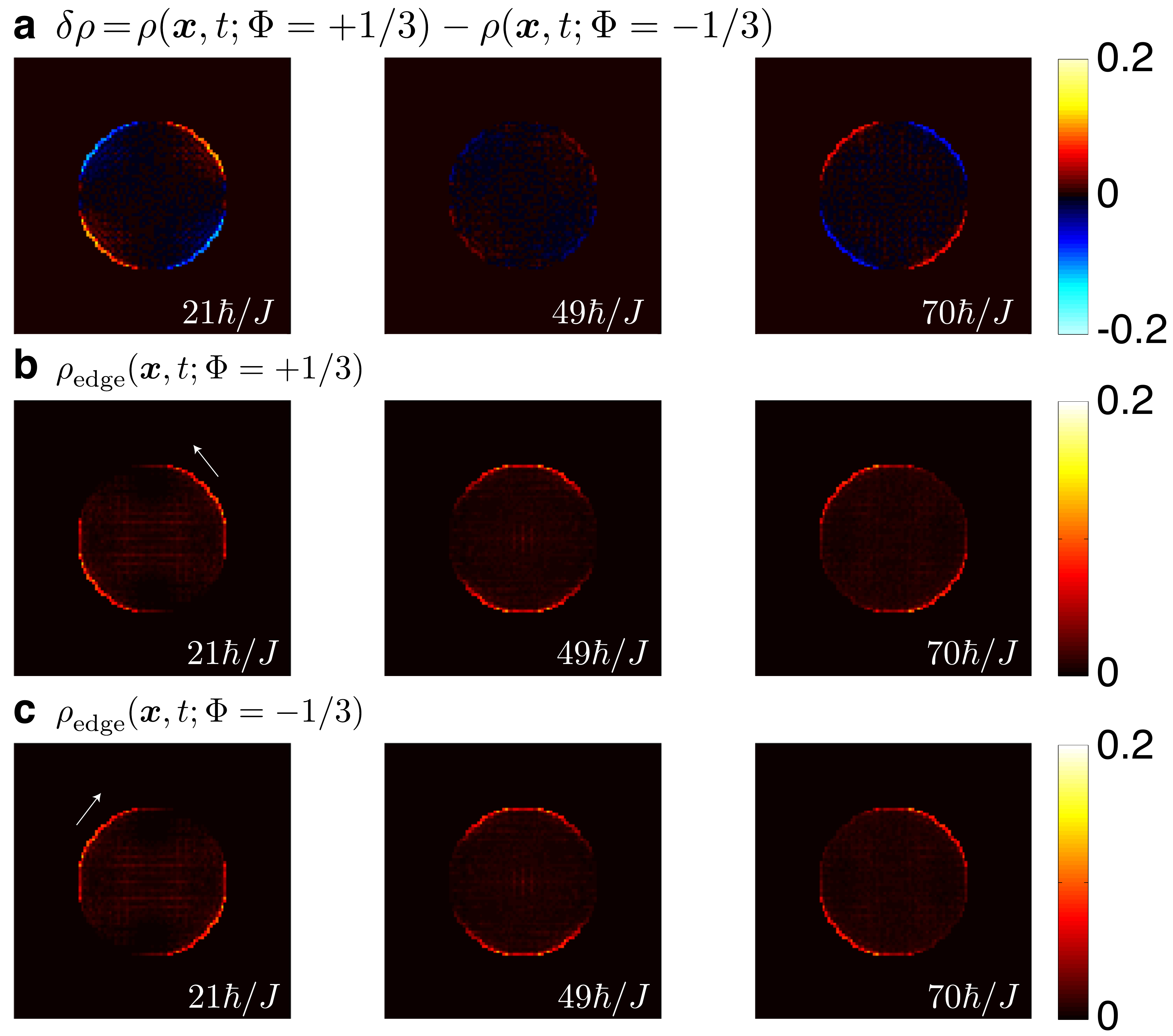}
	\caption{\label{figure5} {\bf The opposite-flux method for dispersive systems.} {\bf a,} Evolution of the difference $\delta \rho =\rho (\bs x , t; \Phi=+1/3)-\rho (\bs x , t; \Phi=-1/3)$, for the same configuration as in Figure \ref{figure3}. This method yields a clear manifestation of the edge states, $\delta \rho \approx \rho_{\text{edge}} (+\Phi)-\rho_{\text{edge}} ( -\Phi)$, by eliminating the undesired contribution of the many bulk states. The edge states chirality is deduced from the evolution of the red and blue patterns. {\bf b,} The edge-states contribution $\rho_{\text{edge}} (\bs x,t)$ for $\Phi=+1/3$, and {\bf c,} for $\Phi=-1/3$.  In the central column, we note the vanishing of the signal $\delta \rho (\bs x) \approx 0$ that occurs at time $t^*\approx 49 \hbar /J$, indicating that the edge states angular velocity is $\dot \theta \approx 0.03 J/\hbar$ for $R_F=27 a$ and $\gamma=\infty$ (also see Appendix~\ref{app:opposite}).}
\end{figure}

We now describe two methods for isolating the edge-states contribution $\rho_{\text{edge}}$ from that of the bulk states, useful for systems with dispersive bulk bands.  The first method consists in performing two successive experiments, using the same apparatus and parameters, but with opposite fluxes $\Phi$ and $- \Phi$.  The difference between the two images $\delta \rho (\bs x , t)=\rho (\bs x , t; +\Phi)-\rho (\bs x , t; -\Phi),$ lacks the non-chiral contribution of the bulk states and is simply given by $\delta \rho \approx \rho_{\text{edge}} (+ \Phi) - \rho_{\text{edge}} (- \Phi)$, see Fig. \ref{figure5}. For our bat geometry, $\rho_{\text{edge}} (+ \Phi) \approx \rho_{\text{edge}} (- \Phi)$ when the edge states have undergone a rotation of $\theta=\pi/2$. This determines the time $t^*= \pi/(2 \dot \theta)$ when the signal $\delta \rho (\bs x, t^*)$ disappears,  giving the angular velocity of the edge states.  This situation is illustrated in Fig.~\ref{figure5} for the $\Phi=1/3$ ``dispersive'' case, where we find $\delta \rho (\bs x, t^*\approx 49 \hbar /J) \approx 0$, in good agreement with the angular velocity $\dot \theta_e \approx 0.03 J/\hbar$ of the populated edge states \cite{Goldman:2012prl,Goldman:2012prlbis}.  We verified that slight differences in the filling [e.g., $E_{\text{F}} (\Phi_+=+ 1/3) \approx E_{\text{F}} (\Phi_-=-1/3) \pm 0.1J$], for example due to finite temperature effects between the two successive experiments, or  variations in the flux (e.g., $\Phi_+=1/3$ and $\Phi_- \approx -\Phi_+ \pm0.01$), do not significantly affect the signal $\delta \rho (\bs x,t)$, highlighting the robustness of this method against experimental imperfections. 

The second method aims to efficiently reduce the bulk dispersion by suddenly \emph{lowering} the  potential walls $V_{\text{hole}}$ at $t=0$, instead of removing them completely. This operation can be achieved in such a way that only the edge states with sufficiently high energies are allowed to propagate, while leaving the bulk states away from the holes. This ``edge-filter method'' can be realized by setting the Fermi energy within the first bulk gap, and then suddenly lowering the  potential $V_{\text{hole}}$ to the value $V_{\text{hole}}^{t>0} \sim W$ at $t=0$, where $W$ is the width of the lowest bulk band. The great efficiency of this method is presented in the Appendix \ref{app:filter} for the case $\Phi=1/3$.

\section{Robustness of the edge states against disorder}

\begin{figure}
	\includegraphics[width=1\columnwidth]{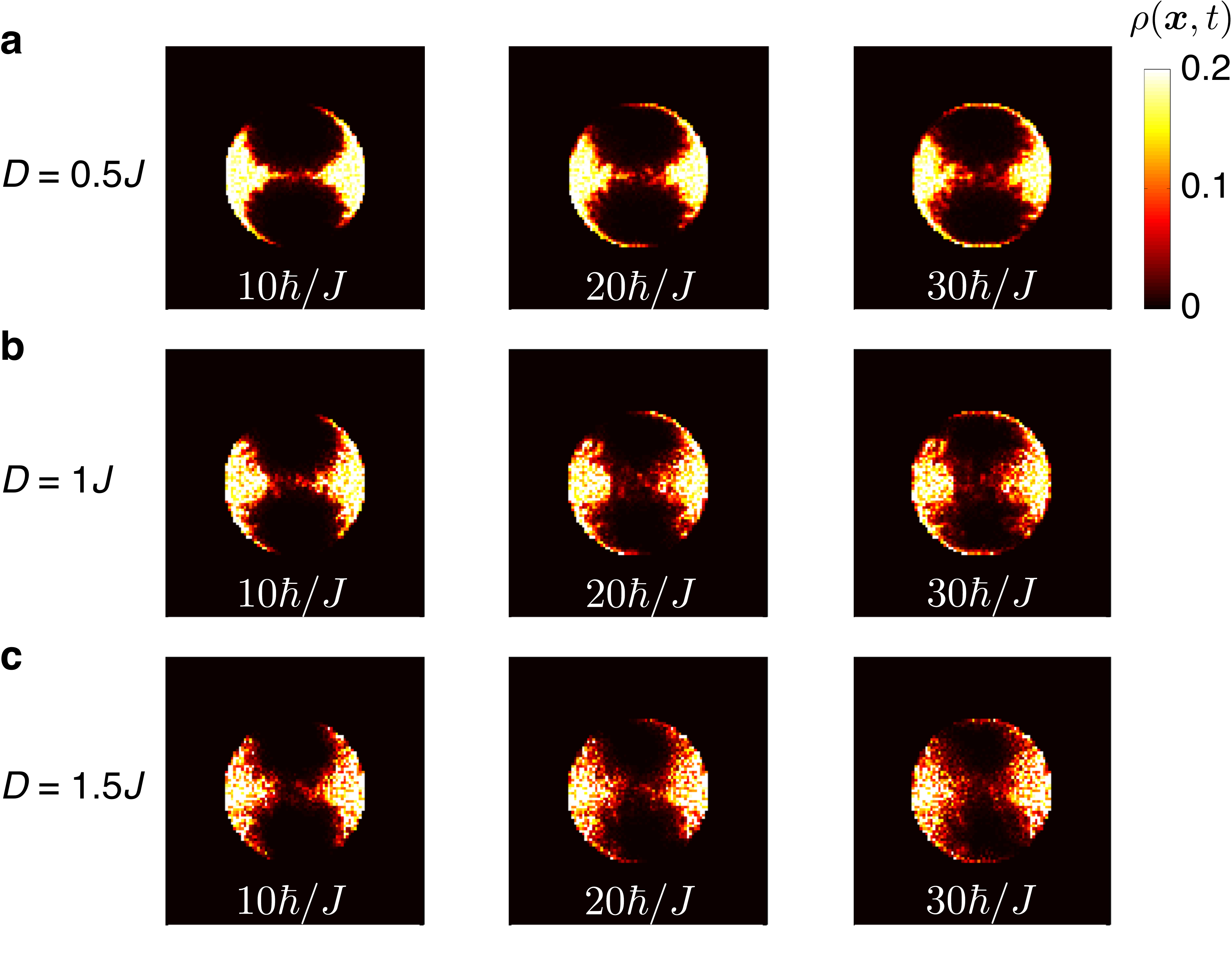}
	\caption{\label{figdisorder}  {\bf Effects of disorder.} The spatial density $\rho (\bs x , t)$ for $\Phi=1/5$, $E_{\text{F}}=-1.5 J$, $r_0=27a$ and $\gamma=\infty$. The disorder strength is {\bf a,} $D=0.5 J$, {\bf b,} $D=1J$ and {\bf c,} $D=1.5J$.}
\end{figure}

We now investigate the robustness of the edge states motion in the presence of disorder. This perturbation, which plays a fundamental role in solid-state systems, can be engineered in optical-lattice setups, e.g., using speckle potentials \cite{Sanchez2010}. We study the effects of disorder by considering a random potential $V_{\text{rand}}$, with energies uniformly distributed within the range $V_{\text{rand}}(\bs x) \in [-D, D]$. The results are presented in Fig. \ref{figdisorder}, for the case $\Phi=1/5$. We find that the chiral edge states signal remains robust for disorder strengths $D \lesssim 1.5 J \approx \Delta$, where $\Delta$ is the size of the bulk gap. Interestingly, we can still distinguish a cyclonic cloud -- a signature of the edge states chirality -- for values up to $D \sim 3 J \approx 2 \Delta$. The immunity of the chiral edge states against disorder, a hallmark of the quantum Hall effect, could thus be demonstrated using our cold-atom setup.

\section{Gaussian walls and smooth circular confinements}

\begin{figure}
	\includegraphics[width=1\columnwidth]{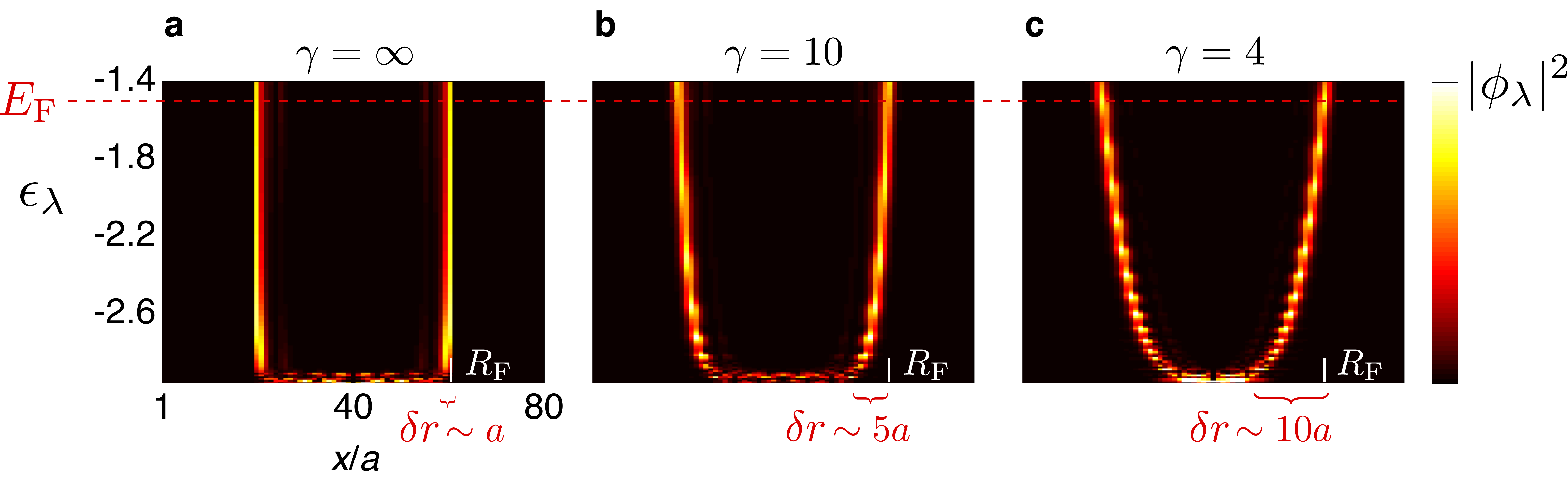}
	\caption{\label{figure10} {\bf Smooth confinements and edge states.}  The amplitudes $\vert \phi_{\lambda}  (x,y)\vert^2$ of the single-particle wavefunctions as a function of the $x$ coordinate and their corresponding energy $\epsilon_{\lambda}$, and setting $y$ at the center of the trap. Here, $\Phi=1/5$, and the external potential is given by $V_{\text{conf}}(r)= V_0 (r/r_0)^{\gamma}$, with $r_0=21 a$ and $V_0=J$. {\bf a,} $\gamma= \infty$, {\bf b,} $\gamma= 10$, and {\bf c,} $\gamma= 4$. The Fermi radius $R_{\text{F}}$ is shown for $E_{\text{F}}=-1.5J$. The populated edge states $\phi_e$ are localized on the radii $R_e=R_e (\epsilon_{e})$, see Eq.~\eqref{radius}, leading to a broadening $\delta r$ of their contribution to the spatial density $\rho_{\text{edge}}$ for finite $\gamma$. }
		\end{figure}

In the absence of walls $V_{\text{hole}}=0$, the edge states lying in the first bulk gap are radially localized, with a radius determined by their energy and the circular confinement. Writing the circular confinement as $V_{\text{conf}}(r)= V_0 (r/r_{0})^{\gamma}$, we find that an edge state $\phi_e$ with energy $\epsilon_e$ is characterized by a localization radius 
\begin{equation}
R_e=r_0\sqrt[\gamma]{\vert \epsilon_e - \epsilon_{\text{min}} \vert/V_0}, \label{radius}
\end{equation}
where $\epsilon_{\text{min}}$ is the minimum of the bulk band. This result is illustrated in Fig. \ref{figure10}, for $r_0=21 a$, $V_0=J$ and $\gamma= \infty, 10, 4$, where the wavefunctions amplitudes $\vert \phi_{\lambda}  (x,y)\vert^2$ are plotted as a function of the $x$ coordinate, and their corresponding energies $\epsilon_{\lambda}$. For an infinitely abrupt trap  \cite{Meyrath2005,Gaunt2012}, $\gamma=\infty$, all the edge states are located at the constant Fermi radius $R_{\text{F}}=r_0$. Therefore, the edge states contribution to the density $\rho_{\text{edge}}$ yields a clear circular signal, with localization length of the order of the lattice spacing $a$. In contrast, for finite $\gamma$, the populated edge states are localized on different radii $R_e \in [R_{\text{F}} - \delta r, R_{\text{F}}]$, leading to a broadening of the edge-state signal $\rho_{\text{edge}}$. For the situation illustrated in Fig. \ref{figure10}, this broadening is of the order $\delta r \sim 5 a$ for $\gamma=10$ and $\delta r \sim 10 a$ for $\gamma=4$. Let us stress another crucial aspect of these smoothly confined QH systems, which is the fact that the angular velocity $\dot{\theta}$ of the chiral edge states, as well as the number $N_{\text{edge}}$ of available edge states within a bulk gap, highly depend on the potential's smoothness $\gamma$: the angular velocity is maximized for highly abrupt confinements ($\gamma \rightarrow \infty$), while the number $N_{\text{edge}}$ is larger for smooth potentials \cite{Goldman:2012prl,Goldman:2012prlbis,StanescuEA2010,Buchhold2012}. We numerically evaluated the angular velocity of the edge states \cite{Goldman:2012prl,Goldman:2012prlbis} for $\Phi=1/5$ and $r_0=21 a$, and we found $\dot \theta_e \approx 0.06 J/\hbar$ for $\gamma=\infty$, $\dot \theta_e \approx 0.02 J/\hbar$ for $\gamma=10$ and $\dot \theta_e < 0.01 J/\hbar$ for $\gamma=4$ (and we note that the angular velocity $\dot \theta_e \propto 1/R_e$). Scaling to a more realistic radius $r_0=100a$, we find that the edge states, which are populated below $E_{\text{F}}=-1.5 J$, undergo a rotation of $\pi/5$ after a time
\begin{align}
&t(\theta=\pi/5) \sim 50 \hbar/J , \quad & &\gamma=\infty \, \, (r_0=100a), \notag  \\
&t(\theta=\pi/5) \sim 150 \hbar/J , \quad & &\gamma=10  \, \,(r_0=100a), \notag \\
&t(\theta=\pi/5) \sim 300 \hbar/J , \quad & &\gamma=4  \, \,(r_0=100a), \notag
\end{align}
indicating that it is highly desirable to design a sharp circular confining trap $\gamma \gg 10$  \cite{Meyrath2005,Gaunt2012}, to clearly observe the edge states rotating motion during reasonable experimental times $t \sim 10-100 \hbar /J$.\\

\begin{figure}[h!]
		\includegraphics[width=1\columnwidth]{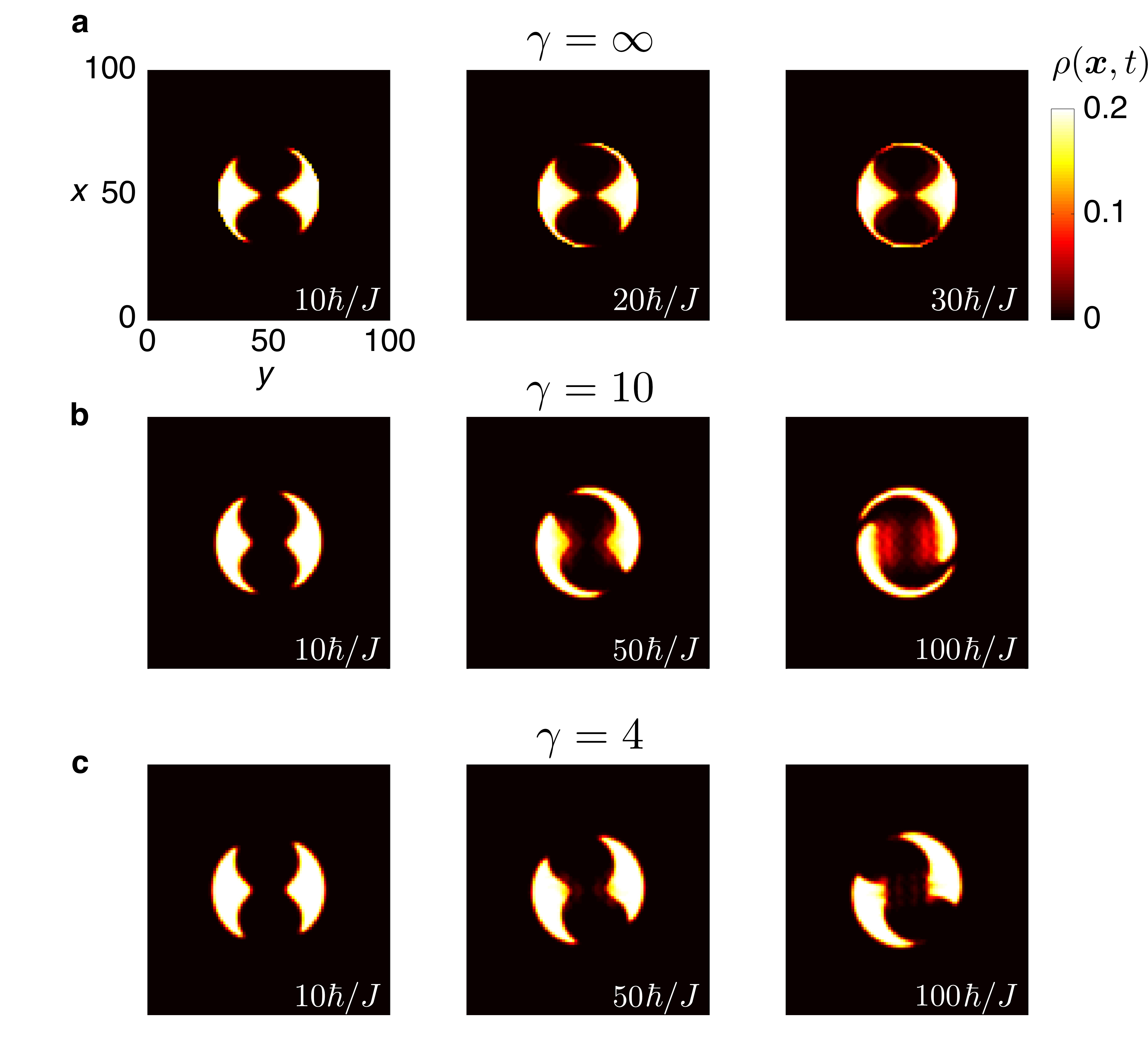}
	\caption{\label{figure11} {\bf Smooth confinements and the density evolution.} Evolution of the spatial density for $\Phi=1/5$ and $E_{\text{F}}=-1.5 J$. The walls $V_{\text{hole}}$ are produced by asymmetric gaussian potentials, with standard deviations $\sigma_y \approx \sqrt{2} \sigma_x$. The external confining potential is $V_{\text{conf}}(r)= V_0 (r/r_0)^{\gamma}$, with $r_0=21 a$ and $V_0=J$.   {\bf a,} $\gamma= \infty$, {\bf b,} $\gamma= 10$, and {\bf c,} $\gamma= 4$. Note the broadening $\delta r$ of the edge-state signal, see Fig. \ref{figure10},  and also, the deceleration of the motion as the confinement becomes smoother $\gamma=\infty \rightarrow 4$.}
\end{figure}

We now investigate the  density evolution $\rho (\bs x,t)$ for smooth confining traps and initial gaussian walls $V_{\text{hole}}$. First of all, we note that the presence of gaussian walls  does not destroy the edge states lying within the first bulk gap. In this bat geometry, the edge states are localized on the outer edge delimited by $V_{\text{conf}} (r)$, but also, on the smooth boundary delimited by the gaussian potentials. Therefore, when $\gamma \sim \infty$, the edge states behave as in Fig. \ref{figure10}{\bf a} in the vicinity of the outer circular edge $r \approx r_0$, whereas they behave similarly as in Fig. \ref{figure10}{\bf c} in the vicinity of the gaussian walls. We point out that, in our scheme, it is the behavior of the edge states near the outer circular edge which plays an important role. Indeed, as shown in Fig. \ref{figure11}{\bf a}, replacing the infinitely abrupt walls $V_{\text{hole}}$ by gaussian potentials does not qualitatively affect the   evolution of the density $\rho (\bs x, t)$ presented in Fig. \ref{figure4}{\bf a}. 

However, as can be anticipated from the discussion above, replacing the perfectly sharp potential $V_{\text{conf}} (r)$  by smoother confinements, $\gamma=\infty \rightarrow 4$, has dramatic consequences on the dynamics. In Fig. \ref{figure11}{\bf b}, which shows the evolution of the density for $\gamma=10$, we clearly observe the broadening $\delta r \sim 5 a$ of the edge-state contribution $\rho_{\text{edge}} (\bs x,t)$, as they progressively encircle the holes. We also note the slower motion undergone by the edge states, which have a reduced angular velocity $\dot \theta (\gamma=10)\sim \dot \theta (\gamma=\infty)/3$, see above. An even more dramatic situation is illustrated for the case $\gamma=4$ in Fig. \ref{figure11}{\bf c}. These results demonstrate the robustness of the edge state motion in the presence of smooth confining traps and gaussian walls, but they greatly emphasize the importance of designing sharp external confinements to improve the experimental detectability of the topological edge states. \\

\section*{{\blue {\Large Conclusions}}}

In this article, we introduced a simple, yet powerful, method to image the dynamics of topological edge states in atomic systems. Our scheme shapes an atomic gas, initially prepared in a topological phase, and directly images its time evolution. By explicitly revealing the presence of propagating chiral edge modes, this method provides an unambiguous signature of topological order in the context of cold atomic gases. Importantly, we have discussed the applicability of our method under realistic experimental conditions, emphasizing the importance of using sharp confining potentials to improve the detection of the edge states signal. The schemes introduced in this work to reduce, or even eliminate, the irrelevant contribution of dispersive bulk states can be applied to a wide family of topological atomic systems, such as the promising Haldane-like optical lattice \cite{alba2011a,Goldman:2012njp} and fractional QH atomic gases \cite{Sorenson:2005,cooper2008a,Grass2012}. Finally, our method can be directly extended to visualize the propagation of $Z_2$ topological (spin-polarized) edge states, both in 2D \cite{Goldman2010a,Beri:2011} and 3D \cite{Bermudez:2010prl}, by using standard spin-dependent imaging methods \cite{Weitenberg2011}.   \\

\section*{{\blue {\Large Acknowledgments} }}

We thank J. Beugnon, I. Bloch, G. Bulnes Cuetara, M. M\"uller, and S. Nascimb\`ene for discussions. This work was supported by the Fonds de la Recherche Scientifique (FNRS Belgium), Agence Nationale de la Recherche (ANR) via the project AGAFON (Artificial gauge fields on neutral atoms), European Research Council (ERC) project Quantum Gauge Theories and Ultracold Atoms (QUAGATUA), the Emergences program (Ville de Paris and Universit\'e Pierre et Marie Curie), and ERC Many-Body physics in gauge fields with ultracold Yb atoms in optical lattices (ManyBo) Starting Grant. Work at Innsbruck is supported by the integrated project AQUTE, the Austrian Science Fund through Spezialforschungsbereich (SFB) F40 FOQUS, and by the Defense Advanced Research Projects Agency (DARPA) Optical Lattice Emulator (OLE) program. I.B.S. acknowledges the financial support of the National Science Foundation (NSF) through the Physics Frontier Center at Joint Quantum Institute (JQI) and the Army Research Office (ARO) with funds from both the Atomtronics Multidisciplinary University Research Initiative (MURI) and DARPA OLE Program.

\newpage

\appendix

\setcounter{figure}{0}
\renewcommand\thefigure{\Alph{figure}}

\section{The time evolution, observable quantities and states population}
\label{app:one}

The system is prepared in the ground-state of the initial Hamiltonian,
\begin{align}
&\hat{H}_0=\hat{H} + \sum_{m,n} V_{\text{hole}}(m,n)\, \hat c^{\dagger}_{m,n} \hat c_{m,n} \label{hamzero}, 
\end{align}
where the potential $V_{\text{hole}}$ describes the walls initially present in the trap (see main text), and where $\hat{H}$ is given by Eq. \eqref{ham}. We denote the number of available sites before and after removing the walls by $n_{\text{sites}}^{0}$ and $n_{\text{sites}}$, respectively, and we define the ratio $(1-\eta)=n_{\text{sites}}^{0}/n_{\text{sites}}$. When $V_{\text{hole}}=0$, the total number of sites within the trap is approximatively given by the area $n_{\text{site}} \approx \pi (r_0/a)^2 $, while the outer circular edge contains about $n_{\text{edge}} \approx 2 \pi r_0/a $ lattice sites. The holes in the density created by $V_{\text{hole}}$ correspond to $n_{\text{hole}}=n_{\text{sites}}-n_{\text{sites}}^{0} \approx \pi (r_0/a)^2/\sqrt{2}$ vacant sites, leading to the large ratio $\eta=n_{\text{hole}}/n_{\text{sites}} \approx 1/\sqrt{2}$. For $r_0=27 a$, the system initially contains $n_{\text{site}}^0  \approx 700$ sites and the number of sites delimiting the edge of the bat is  $n_{\text{edge}}^0 \approx 320$ sites. Thus, the spacious holes used in our calculations lead to a large edge/bulk ratio. For $\Phi \approx p/q \in \mathbb{Q}$, and initially setting the Fermi energy in the lowest bulk gap, leads to the filling factor $\nu^{0} = N_{\text{part}}/ n_{\text{sites}}^{0} \sim 1/q$. After removing the walls $V_{\text{hole}}$, the filling factor is reduced to the smaller value $\nu = N_{\text{part}}/ n_{\text{sites}} \sim (1- \eta)/q \ll \nu^{0}$. \\

The groundstate of Hamiltonian \eqref{hamzero} is written as
\begin{equation}
\vert \Psi_{0} \rangle = \prod_{E_{\alpha} < E_{\text{F}}} \hat{f}^{\dagger}_{\alpha} \vert \emptyset \rangle , \label{groundstate}
\end{equation} 
where the operator $\hat{f}_{\alpha}^{\dagger}$ creates a particle in the single-particle state $\vert \chi_{\alpha} \rangle$, with energy $E_{\alpha}$ located below the Fermi energy $E_{\text{F}}$. Here $\{ \vert \chi_{\alpha} \rangle , E_{\alpha} \}$ represents the complete set of  single-particle eigenstates and eigenvalues satisfying the stationary Schr\"odinger equation 
\begin{equation}
\hat{H}_0  \vert \chi_{\alpha} \rangle = E_{\alpha} \vert \chi_{\alpha} \rangle .
\end{equation}
We are interested in the time evolution of the spatial density $\rho (\bs{x},t)$ after removing the walls $V_{\text{hole}}$ at $t=0$. The evolution of the single-particle states $\vert \chi_{\alpha} \rangle$ is then entirely governed by the Hamiltonian $\hat{H}$. It is therefore convenient to introduce the eigenstates and eigenvalues  $\{ \vert \phi_{\lambda} \rangle , \epsilon_{\lambda} \}$ corresponding to the Hamiltonian $\hat{H}$,
\begin{equation}
\hat{H} \vert \phi_{\lambda} \rangle = \epsilon_{\lambda} \vert \phi_{\lambda} \rangle.
\end{equation} 
We then define $\vert \chi_{\alpha} (t) \rangle$ as the time evolution of the initial state $\vert \chi_{\alpha} \rangle$ according to the Hamiltonian $\hat{H}$, 
\begin{equation}
\vert \chi_{\alpha} (t) \rangle = \sum_{\lambda} \langle \phi_{\lambda} \vert \chi_{\alpha} \rangle e^{-i \epsilon_{\lambda} t / \hbar} \vert \phi_{\lambda} \rangle .\label{statevolution}
\end{equation}
The spatial density $\rho (\bs{x},t)$ at time $t$ is given by
 \begin{equation}
\rho (\bs{x},t) = \sum_{E_{\alpha} < E_{\text{F}}} \vert  \chi_{\alpha} (\bs{x}, t) \vert ^2,
\end{equation}
namely, the particle density $\rho (\bs{x},t)$ is entirely governed by the time-evolution of the initially occupied single-particle states. The time evolution of the atomic cloud, after releasing the walls $V_{\text{hole}}$ at $t=0$, can therefore be numerically evaluated through a direct diagonalization of the Hamiltonians $\hat{H}$ and $\hat{H}_0$. In our study, a crucial aspect consists in identifying the regimes for which the edge states propagating around the initially forbidden regions provide a clear signal, which is not perturbed by the many bulk states. It is therefore desirable to separately evaluate the contributions of the initially populated bulk and edge states. We introduce the corresponding quantities 
\begin{align}
&\rho_{\text{edge}} (\bs{x},t) = \sum_{E_{e}< E_{\text{F}}} \vert  \chi_{e} (\bs{x}, t) \vert ^2, \notag \\
&\rho_{\text{bulk}} (\bs{x},t) = \rho (\bs{x},t) -\rho_{\text{edge}} (\bs{x},t), \notag 
\end{align}
where the sum $\sum_{E_{e}< E_{\text{F}}}$ is restrained to the populated edge states with energies $E_e$ located within the bulk gap.\\

\begin{figure}[h!]
	\centering
	\includegraphics[width=1\columnwidth]{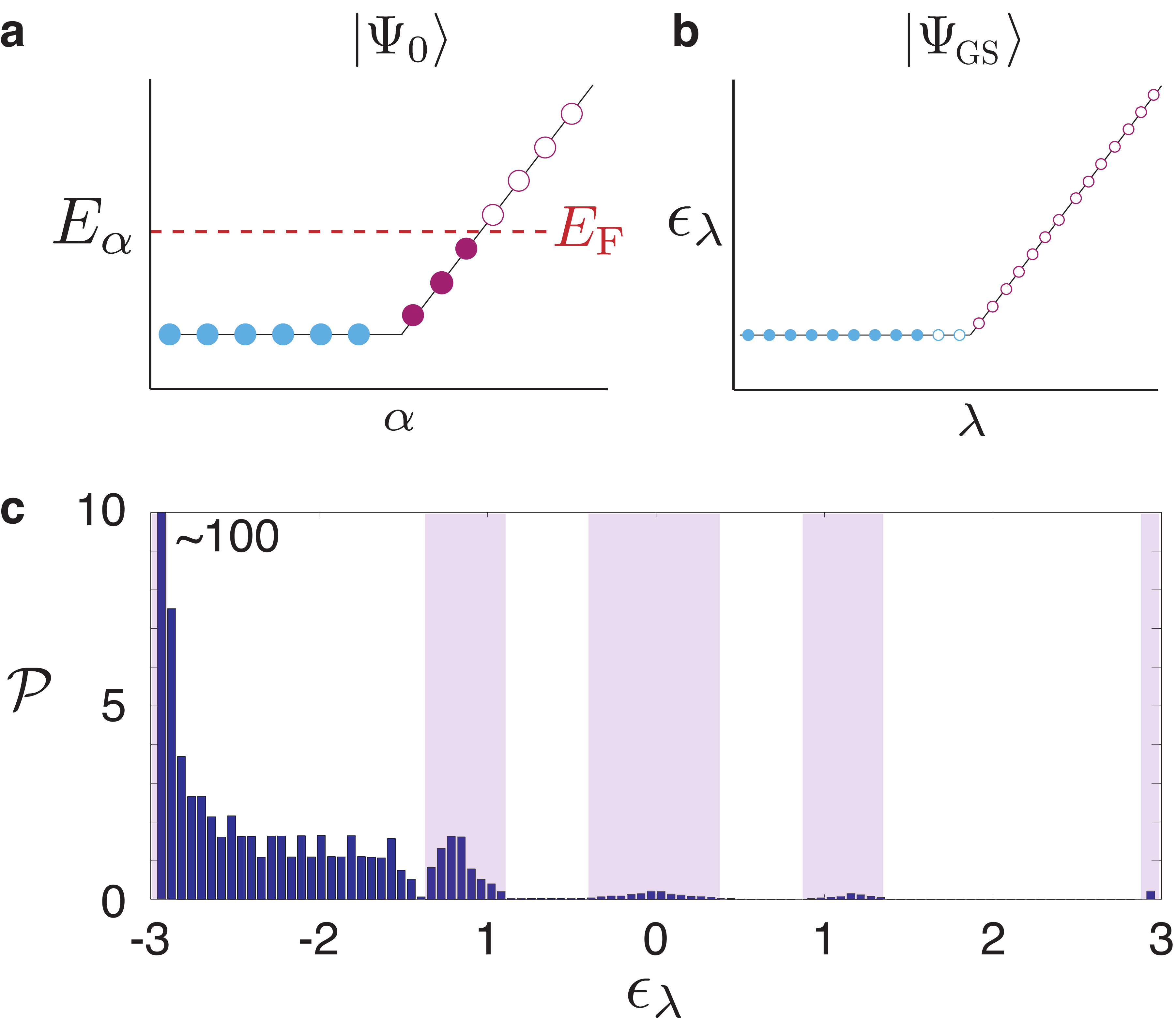}
	\caption{\label{figure6} {\bf Ground states and single-particle states population.} Comparing {\bf a,} the groundstate  $\vert \Psi_{0} \rangle$ of the initially prepared system (with holes) and {\bf b,} the groundstate $\vert \Psi_{\text{GS}} \rangle$ of the final system (without holes). Filled and empty blue (red) dots represent the occupied and unoccupied bulk (edge) states, respectively, with occupation number $=1$. Note that the total number of particles, $N_{\text{part}}=\sum_{E_{\alpha} < E_{\text{F}}}$, is constant. {\bf c} The population $\mathcal{P}$ of the states $\vert \phi_{\lambda} \rangle$, as a function of their energy $\epsilon_{\lambda}$, as established by the Fermi energy $E_{\text{F}}=-1.5 J$ for the case illustrated in Fig. \ref{figure4} (i.e., $\Phi=1/5$, $\gamma=\infty$, $r_0=27 a$ and $\eta \approx 0.7$). The energies corresponding to bulk states are emphasized by purple shaded regions, see Fig. \ref{figure2}{\bf b}.}
	\end{figure}

\begin{figure}[h!]
	\centering
	\includegraphics[width=1\columnwidth]{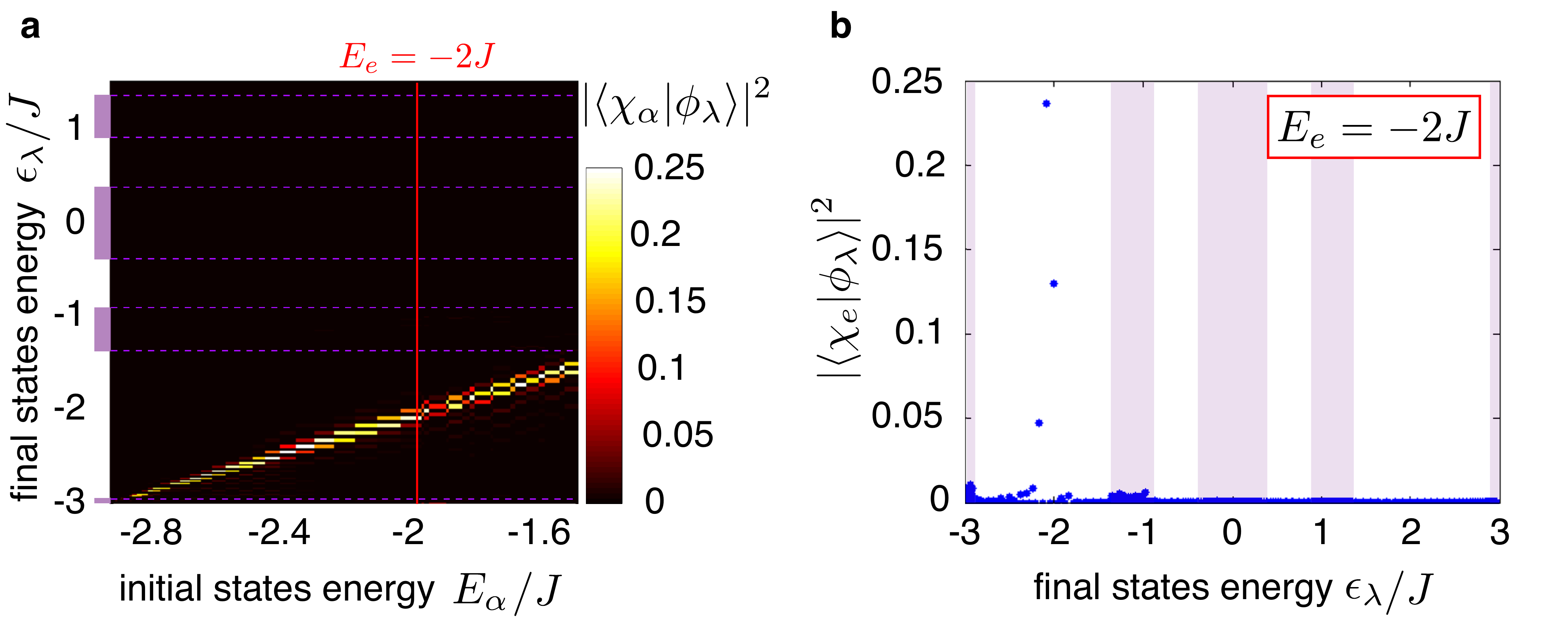}
	\caption{\label{figure6bis} {\bf Wave-function overlaps between initial and final states.} {\bf a,} The overlap $\vert \langle \chi_{\alpha}  \vert \phi_{\lambda} \rangle \vert ^2$ between the eigenstates of the initial ($\chi_{\alpha}$) and final ($\phi_{\lambda}$) Hamiltonians, represented as a function of their energies. The parameters are the same as in Fig. \ref{figure6}. Note that the edge states are found in the bulk gap, namely, within the range $E \approx [-2.9 J, -1.3 J]$. {\bf b,} Cut through the plot in Fig. {\bf a}, for a specific initial edge state with energy $E_e=-2 J$. The initial edge modes effectively project to the final edge modes, highly limiting the population of high energy (dispersive) bulk states. The energies corresponding to bulk states are emphasized by purple shaded regions in {\bf a} and {\bf b}.}
\end{figure}

In this study, the time evolution is chosen to be entirely dictated by the Hamiltonian $\hat{H}$, and thus, it is non-dissipative: after releasing the walls, the total energy of the system is constant and is given by
\begin{equation}
\mathcal{E}_0= \langle  \Psi_{0} \vert \hat{H} \vert \Psi_{0} \rangle \approx \sum_{E_{\alpha} < E_{\text{F}}} E_{\alpha},
\end{equation}
where we considered the approximation $\langle  \Psi_{0} \vert \hat{H} \vert \Psi_{0} \rangle \approx \langle  \Psi_{0} \vert \hat{H}_0 \vert \Psi_{0} \rangle$, which is valid for $V_{\text{hole}}$ sufficiently abrupt. For example, in the situation illustrated in Fig. \ref{figure4}, the energy released after removing the walls is about $J/3$. Therefore, in our calculations, the many-body state $\vert \Psi (t) \rangle$ never reaches the ground state $\vert \Psi_{\text{GS}} \rangle$ of the final Hamiltonian  $\hat{H}$, with energy
\begin{equation}
\mathcal{E}_{\text{GS}}= \langle  \Psi_{\text{GS}} \vert \hat{H} \vert \Psi_{\text{GS}} \rangle .
\end{equation}
This final ground state, which differs from the initially prepared ground state $\vert \Psi_{0} \rangle$ in Eq. \eqref{groundstate}, is expressed as 
\begin{equation}
\vert \Psi_{\text{GS}} \rangle= \prod_{\lambda =1}^{N_{\text{part}}} \hat{a}^{\dagger}_{\lambda} \vert \emptyset \rangle ,
\end{equation} 
where $\hat{a}^{\dagger}_{\lambda}$ creates a particle in the single-particle state $\vert \phi_{\lambda} \rangle$, with energy $\epsilon_{\lambda}<\epsilon_{\lambda +1}$. Here, $N_{\text{part}}=\sum_{E_{\alpha} < E_{\text{F}}}$ is the total number of particles in the system, which is supposed to be constant (see Figs. \ref{figure6}{\bf a}-{\bf b}).  In our non-dissipative framework, the probability $\mathcal{P}_{\lambda} (t)$ of finding a particle in the eigenstate $\vert \phi_{\lambda} \rangle$  is constant and inferior to one, as it is simply given by
\begin{align}
\mathcal{P}_{\lambda} (t)&= \langle \Psi (t) \vert \hat a^{\dagger}_{\lambda} \hat a_{\lambda} \vert \Psi (t) \rangle= \sum_{E_{\alpha} < E_{\text{F}}} \vert  \langle \phi_{\lambda} \vert \chi_{\alpha} \rangle \vert ^2 = \text{cst} < 1, \notag \\
&\ne \langle \Psi_{\text{GS}} \vert \hat a^{\dagger}_{\lambda} \hat a_{\lambda} \vert \Psi_{\text{GS}} \rangle = 1 \text{ if $\lambda \le N_{\text{part}}$ (\text{otherwise} 0)}. \label{poplambda}
\end{align}
The populations $\mathcal{P}_{\lambda}$ are illustrated as a histogram in Fig. \ref{figure6}{\bf c}, for the case $\Phi=1/5$ and $E_{\text{F}}=-1.5 J$. We find that when the Fermi energy is initially set within the first bulk gap, the population of high-energy dispersive bulk bands is highly limited during the evolution, which guarantees a clear spatial separation of the bulk and edge signal in this ``topological quasi-flat band'' configuration. The efficiency with which the initial edge modes $\chi_{e}$ project to the final edge modes $\phi_e$ is further shown in Fig. \ref{figure6bis}, indicating the success of our scheme. For the situation illustrated in Figs. ~\ref{figure6}-\ref{figure6bis}, we find that the number of populated edge states $\mathcal{P}_{\text{edge}}=\sum_{\lambda \in \text{edge states}}\mathcal{P}_{\lambda} \approx 30 \approx N_{\text{edge}}/2$, where $N_{\text{edge}} \approx 64$ is the number of populated edge states before removing the walls (see main text). This result is in agreement with the fact that the edge delimited by the bat contains $n_{\text{edge}}^0 \approx 2 \times n_{\text{edge}}$ sites, where $n_{\text{edge}} \approx 2 \pi r_0/a$ is the number of sites delimited by the external circular confinement. \\ 

In the main text, we  discuss the non-dissipative evolution of the atomic cloud after releasing the walls $V_{\text{hole}}$, neglecting thermalization processes. However, it is instructive to estimate the energy loss that would be required to reach the ground state $\vert \Psi_{\text{GS}} \rangle$, namely $\delta \mathcal{E}=\mathcal{E}_0 - \mathcal{E}_{\text{GS}}$.   We propose to evaluate this energy difference for a configuration which is particularly relevant for our work (see main text), namely, a system exhibiting large initial holes  and a flat (dispersionless) lowest bulk band. In the following, we set the energy of the lowest bulk band equal to zero. The energy difference $\delta \mathcal{E}$ will generally be large in this ``large hole/flat band'' situation, since many bulk states become available in the lowest energy band after removing the walls, namely $\mathcal{E}_{\text{GS}} \sim 0$ [see Figs. \ref{figure6}{\bf a}-{\bf b} and below]. For this configuration, which is particularly suitable for visualizing the edge states encircling the holes, the energy difference is thus approximatively given by
\begin{equation}
\delta \mathcal{E} \approx \sum_{\text{occupied edge states } e} E_{e}, \label{deltaen}
\end{equation}
where we considered the Fermi energy to be located inside the first bulk gap [see Fig. \ref{figure6}{\bf a}-{\bf b}]. For large holes $\eta=n_{\text{hole}}/n_{\text{sites}} \gg 0$, the final filling factor $\nu \ll 1/q$ such that the lowest bulk band will only be partially occupied when reaching the ground state $ \vert \Psi_{\text{GS}} \rangle$, and thus $\mathcal{E}_{\text{GS}} \sim 0$. From a rough geometric argument, we expect about $N_{\text{edge}} \sim \pi R_{F} (2 + \sqrt{\eta})/a(q-1)$ available edge states in the lowest bulk gap of the initial system. Approximating the edge-state branch as being linear inside the whole gap $\Delta$, we find
\begin{equation}
\delta \mathcal{E} \sim \frac{\Delta \pi}{q-1} (R_F/a) (1+0.5 \sqrt{\eta}), \label{deltaen2}
\end{equation}
which corresponds to an energy difference $\delta \mathcal{E} \sim 1.5 (R_F/a) J$ in the case $\Phi=1/5$ and $\eta = 1/\sqrt{2}$ (see main text). In a typical experiment, $R_F \sim 100 a$, which would require an important energy loss $\delta \mathcal{E} \sim 150 J$ to reach the ground state $\vert \Psi_{\text{GS}} \rangle$. We verified that the estimated energy difference $\delta \mathcal{E}$ in Eq. \eqref{deltaen2} is in good agreement with a direct numerical evaluation of Eq. \eqref{deltaen}, for the system illustrated in Fig. \ref{figure4} (i.e., $\Phi=1/5$, $\eta \approx 1/\sqrt{2}$ and $E_{\text{F}}=-1.5 J$). As illustrated in Fig. \ref{figure6}{\bf b}, the ground state $\vert \Psi _{\text{GS}} \rangle$ generally consists in a partially occupied bulk band, which indicates that  the edge states will no longer be populated if the system relaxes to the ground state $\vert \Psi _{\text{GS}} \rangle$. Therefore, our scheme requires  that the system remains in an excited state during the time evolution, namely, that dissipation should be limited.\\

\section{The opposite-flux method for dispersive bulk bands}
\label{app:opposite}

\begin{figure}[h!]
	\centering
	\includegraphics[width=1.05\columnwidth]{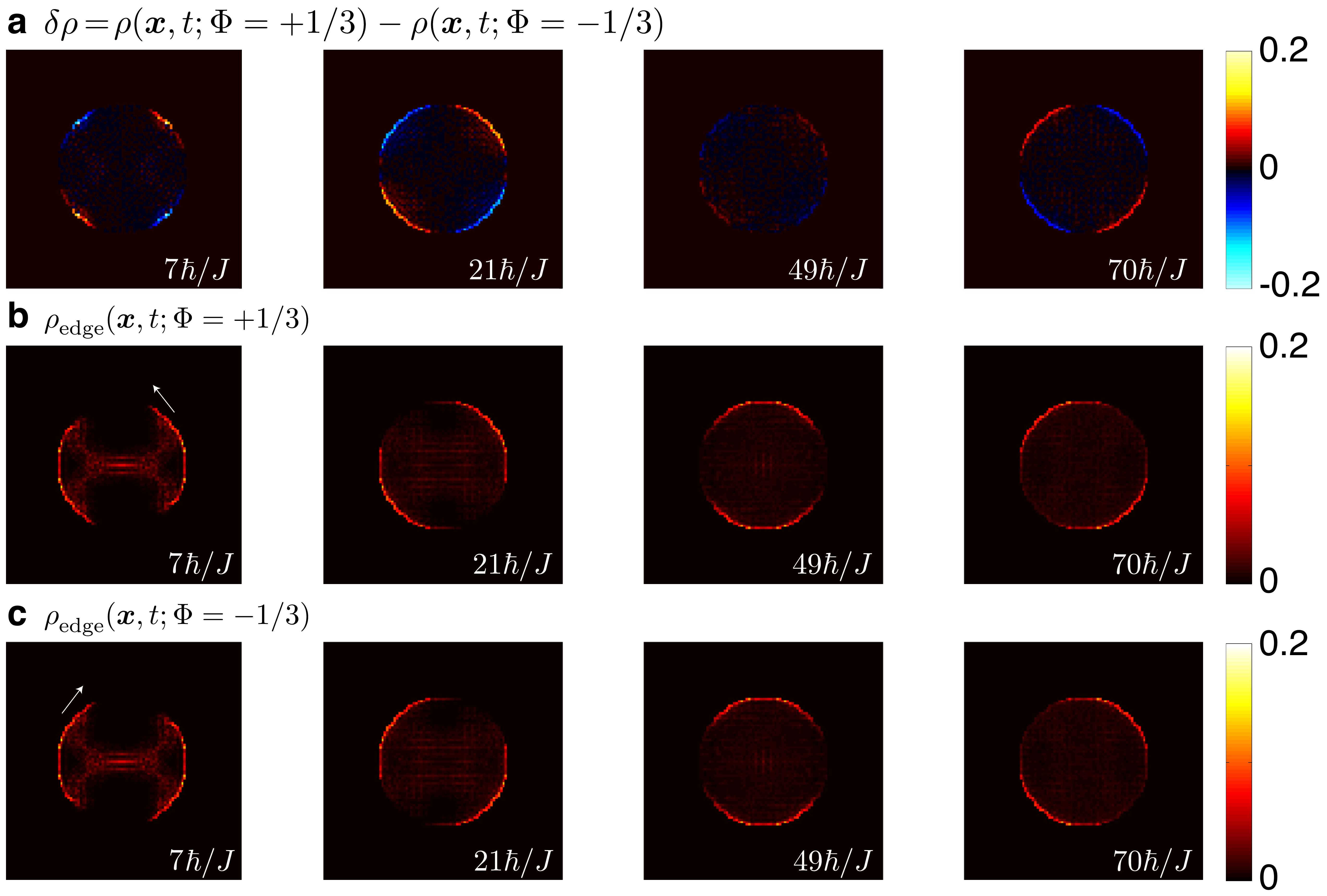}
	\caption{\label{figure7} {\bf The opposite-flux method for dispersive systems.} {\bf a,} Evolution of the difference $\delta \rho =\rho (\bs x , t; \Phi=+1/3)-\rho (\bs x , t; \Phi=-1/3)$, for the same configuration as in Figure 3. {\bf b,} Evolution of the initially populated edge states $\rho_{\text{edge}} (\bs x, t)$ for $\Phi=+1/3$, and  {\bf c,} for  $\Phi=-1/3$. The vanishing of the signal $\delta \rho (\bs x) \approx 0$ corresponds to a rotation of $\pi/2$. This happens at time $t^*\approx 49 \hbar /J$, indicating that the edge state angular velocity is $\dot \theta \sim 0.03 J/\hbar$ for $R_F=27 a$
.}
\end{figure}

In this Section, we discuss how the difference $\delta \rho (\bs x,t) =\rho (\bs x , t; +\Phi)-\rho (\bs x , t; - \Phi)$, obtained from two successive measurements with opposite magnetic fluxes, is related to the edge states contributions $\rho_{\text{edge}} (\bs x, t; \pm \Phi)$. We show in Fig. \ref{figure7}{\bf a}, the evolution of $\delta \rho (\bs x,t)$ for the dispersive case $\Phi=1/3$, which clearly indicates that the contributions from the bulk $\rho_{\text{bulk}} (\bs x, t; + \Phi) \approx \rho_{\text{bulk}} (\bs x, t; - \Phi)$ vanish from the signal at all times $t=7-70 \hbar/J$. Accordingly,  $\delta \rho (\bs x,t) \approx \rho_{\text{edge}} (\bs x, t; + \Phi)-\rho_{\text{edge}} (\bs x, t; - \Phi)$. To clarify the evolution of this signal, which has non-vanishing values in the vicinity of the Fermi radius $R_F$, we show the chiral evolution of the edge states contribution $\rho_{\text{edge}} (\bs x, t; \pm \Phi)$ in Figs. \ref{figure7}{\bf b}-{\bf c}. At small times, the overlap between the two contributions $\rho_{\text{edge}} (\bs x, t; \pm \Phi)$ decreases in time, leading to a progressive broadening of the signal $\delta \rho (\bs x,t)$ along the 1D circular edge. Then, after reaching a rotation of $\theta \approx \pi/4$, the overlap increases, and eventually leads to a vanishing of the signal $\delta \rho (\bs x,t)\approx 0$ when the edge states have undergone a rotation of $\pi/2$, where $\rho_{\text{edge}} (\bs x, t; + \Phi) \approx \rho_{\text{edge}} (\bs x, t; - \Phi)$. In Fig. \ref{figure7}, this happens at time $t^*\approx 49 \hbar /J$, indicating that the edge states angular velocity is $\dot \theta \approx 0.03 J/\hbar$ for $R_F=27 a$. The opposite-flux method therefore offers a general technique for emphasizing the existence of chiral  edge states in dispersive systems, and also, for evaluating their characteristic angular velocity. 

We verified that a slight difference in the filling, $E_{\text{F}} (\Phi_+=+ 1/3) \approx E_{\text{F}} (\Phi_-=-1/3) \pm 0.1J$, or variations in the flux, $\Phi_+=1/3$ and $\Phi_- \approx - \Phi_+ \pm0.01$, does not significantly affect the signal $\delta \rho (\bs x,t)$ shown in Fig. \ref{figure7} {\bf a}, highlighting the robustness of this method against possible experimental imperfections. \\

\section{The edge-filter method for dispersive bulk bands}
\label{app:filter}

\begin{figure}[h!]
	\centering
	\includegraphics[width=1\columnwidth]{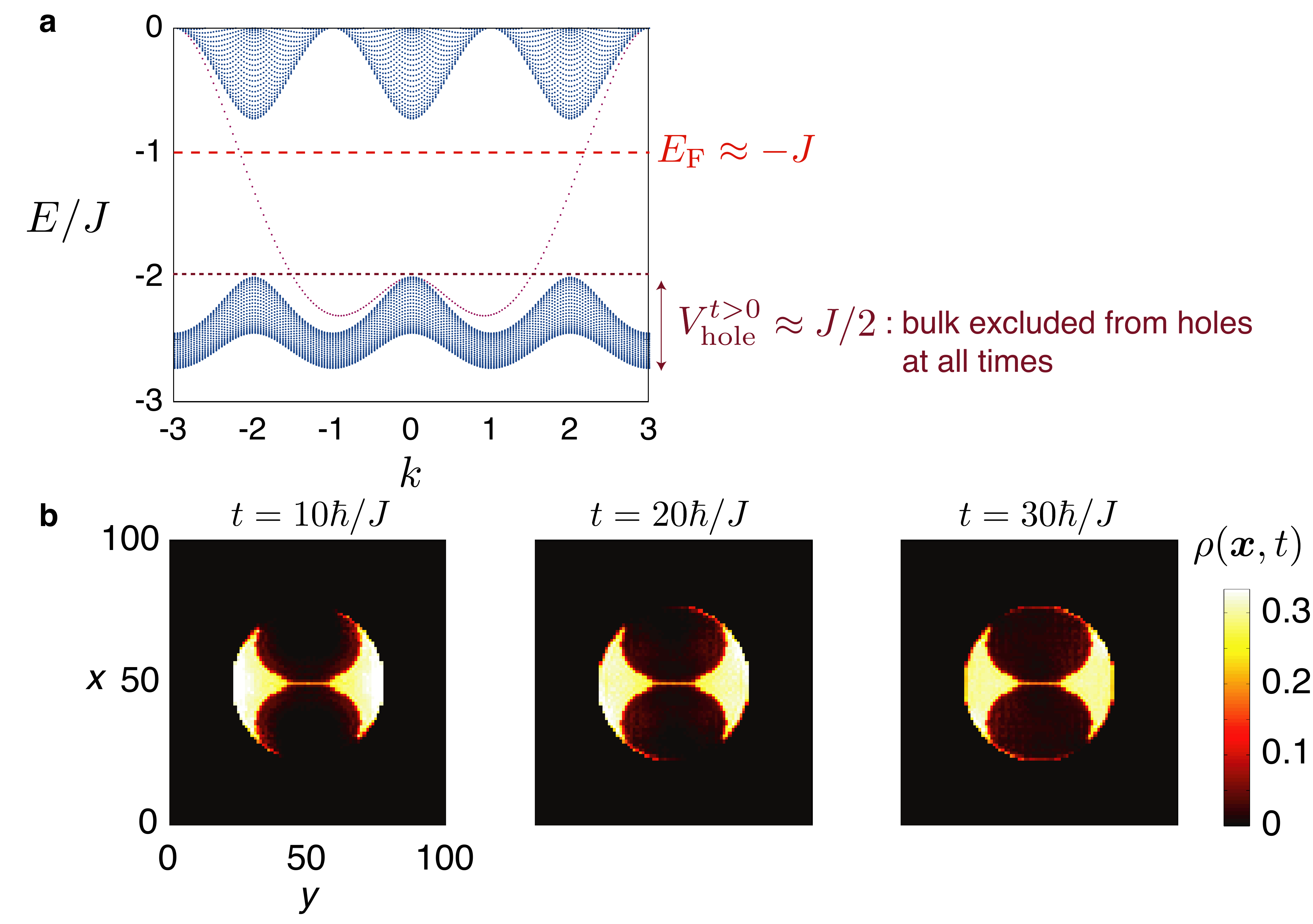}
	\caption{\label{figure8} {\bf The edge-filter method for dispersive systems.} {\bf a,} Energy spectrum $E(k)$ as a function of the quasi-momentum $k$ for $\Phi=1/3$. Also shown are the typical Fermi energy and the final walls potential strength $V_{\text{hole}}^{t>0} \approx W$, where $W$ is the width of the lowest bulk band. {\bf b,} Evolution of the spatial density $\rho (\bs x , t)$ for $\Phi=1/3$, $E_{\text{F}}=- 0.85J$, and infinitely abrupt confinement/walls. At $t=0$, the strength of the walls potential is suddenly reduced to $V_{\text{hole}}^{t>0}=0.5 J$. Note that most of the bulk states are excluded from the holes during the evolution, yielding a clear edge-state signal, to be compared with Fig. \ref{figure3} {\bf a}.}
\end{figure}

Another strategy consists in allowing the edge states to propagate at $t>0$, while \emph{forbidding} the bulk states to penetrate the regions initially occupied by the holes. This can be achieved by suddenly lowering the additional  potential walls $V_{\text{hole}}$ to some intermediate value $V_{\text{hole}}^{t>0}\ne0$, instead of removing them completely at $t=0$. This ``edge-filter'' scheme is illustrated in Fig. \ref{figure8}{\bf a}, for the dispersive case $\Phi=1/3$. By suddenly lowering the walls potential to the value $V_{\text{hole}}^{t>0} \approx W$, where $W$ is the width of the lowest bulk band, we limit the undesired filling of the holes by the bulk states at times $t>0$. In contrast, the populated edge states with energy $\epsilon_e > V_{\text{hole}}^{t>0} + \epsilon_{\text{min}}$, where $\epsilon_{\text{min}}$ is the minimum of the bulk band, are allowed to propagate around the holes without being spoiled by the bulk. The resulting time evolution of the density $\rho (\bs x , t)$, presented in Fig. \ref{figure8}{\bf b}, shows a clear propagation of the edge states around the holes. Experimentally, this method offers an efficient method to isolate the edge states contribution from the spoiling bulk background, but it necessitates a very precise control over the potential strength $V_{\text{hole}}$.

\section{Sensitivity to imperfect filling}
\label{app:sensitivity}

\begin{figure}[h!]
	\centering
	\includegraphics[width=1\columnwidth]{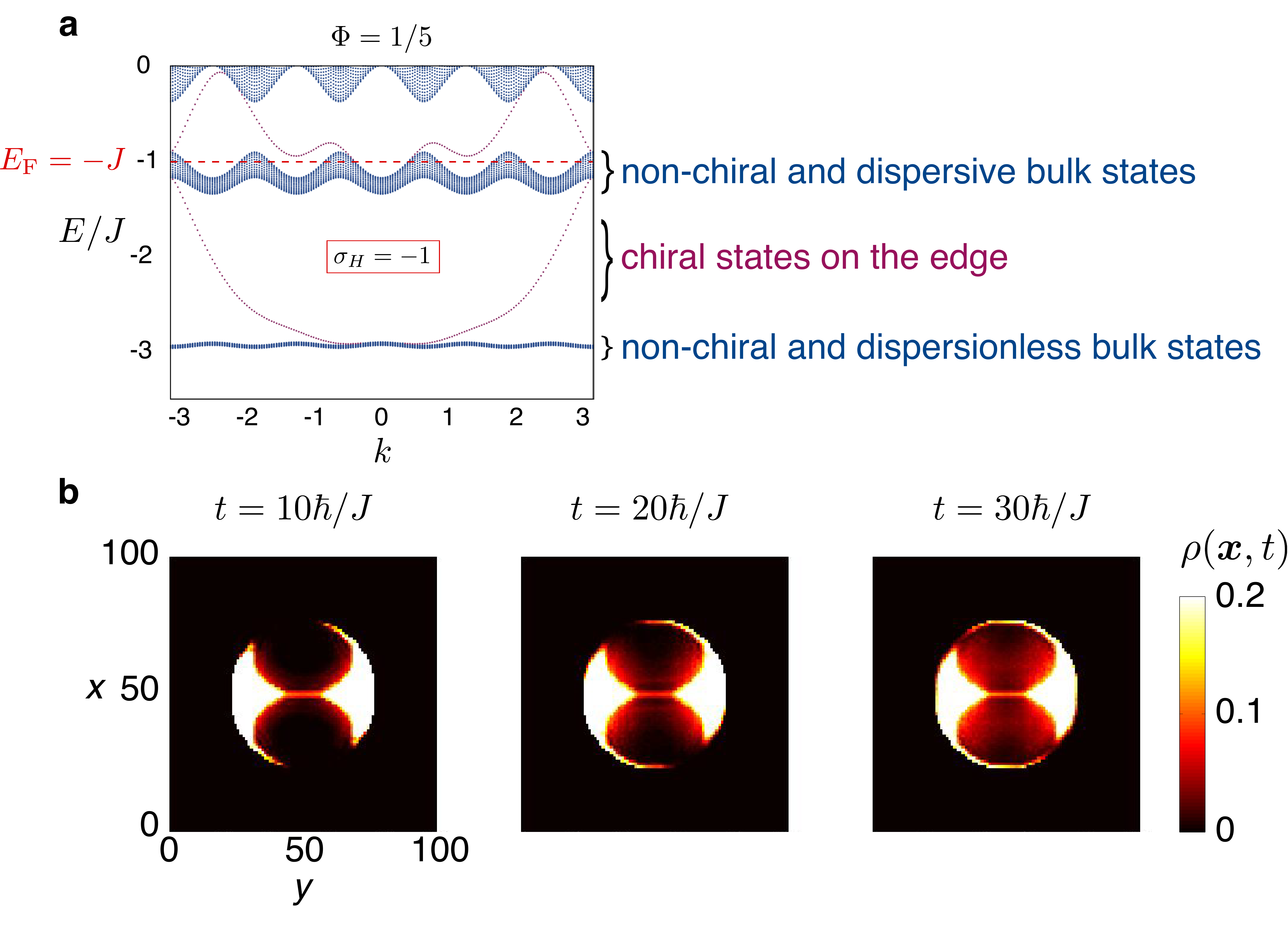}
	\caption{\label{figure9} {\bf Sensitivity to imperfect filling in the flat-band configuration.} {\bf a,} Energy spectrum $E(k)$ as a function of the quasi-momentum $k$ for $\Phi=1/5$, indicating the ``dramatic'' Fermi level $E_{\text{F}}=-J$ used in {\bf b}, as well as the characteristics of the occupied states. {\bf b,} Evolution of the spatial density $\rho (\bs x , t)$ for $E_{\text{F}}=- J$ and infinitely abrupt confinement/walls. Note the non-chiral behavior of the dispersive bulk states, to be compared with Fig. \ref{figure4} {\bf a}.}
\end{figure}

Our general scheme is based on the possibility of preparing a QH atomic state, which can be achieved by generating a magnetic flux $\Phi$ in the lattice and filling the lowest bulk band completely. In other words, one has to tune the total number of particles $N_{\text{part}}$ in such a way that the Fermi energy $E_{\text{F}}$ lies within the lowest bulk gap. According to the topological nature of the lowest bulk band, one is then guaranteed that topological edge states are populated. In practice, the total number of particles (and the corresponding Fermi energy $E_{\text{F}}$) can be tuned with a great precision in cold-atom experiments. However, it is instructive to test the robustness of our method against inexact filling effects, in particular, for the dispersionless case $\Phi=1/5$. We remind that in this configuration, the clear separation of the bulk and edge states contributions to the density relies on the fact that the bulk states are dispersionless (they are described by a quasi-flat band).  Here, we show that this picture holds even when the second (dispersive) bulk band is dramatically filled, see Fig. \ref{figure9}. From this result, we find that the high-energy bulk states contribute in a non-chiral manner to the density evolution, see Fig. \ref{figure9}, and that their dispersive motion is slow compared to the edge states propagation along the circular boundary. In particular, this shows that the chiral picture drawn by the density $\rho (\bs x,t)$, and which will be imaged in an experiment, can unambiguously be attributed to the populated edge states. We conclude that the edge-state signal obtained for the interesting case $\Phi=1/5$ remains clear and detectable, as long as sufficiently many edge states are initially populated. In particular, this indicates that our scheme is robust at finite temperature $T>0$, as long as it remains small compared to the gap's width $\Delta$, in order to insure a sufficiently large edge-state population.

\section{The holes geometry and size effects}
\label{app:size}

Our setup illustrated in Fig. \ref{figure1} features two large repulsive potentials, which create the initial bat geometry. These holes are chosen to be created by infinitely abrupt walls $V_{\text{hole}}$, which are delimited by the two ellipses $$(x \pm r_0/2)^2 + (y/\sqrt{2})^2 =  r_0^2/4, $$ where the coordinates $(x,y)=0$ at the center of the trap. This choice is motivated by the fact  that these walls coincide (up to first order) with the external circular wall $V_{\text{conf}} (r)$ of radius $r_0$, in the vicinity of the poles located at $(\pm r_0) \bs{1}_{\bs{x}}$. Note that in the following of this discussion, we consider that $V_{\text{conf}} (r)\propto (r/r_0)^{\gamma}$ with $\gamma=\infty$. 

\begin{figure}[h!]
	\centering
	\includegraphics[width=1\columnwidth]{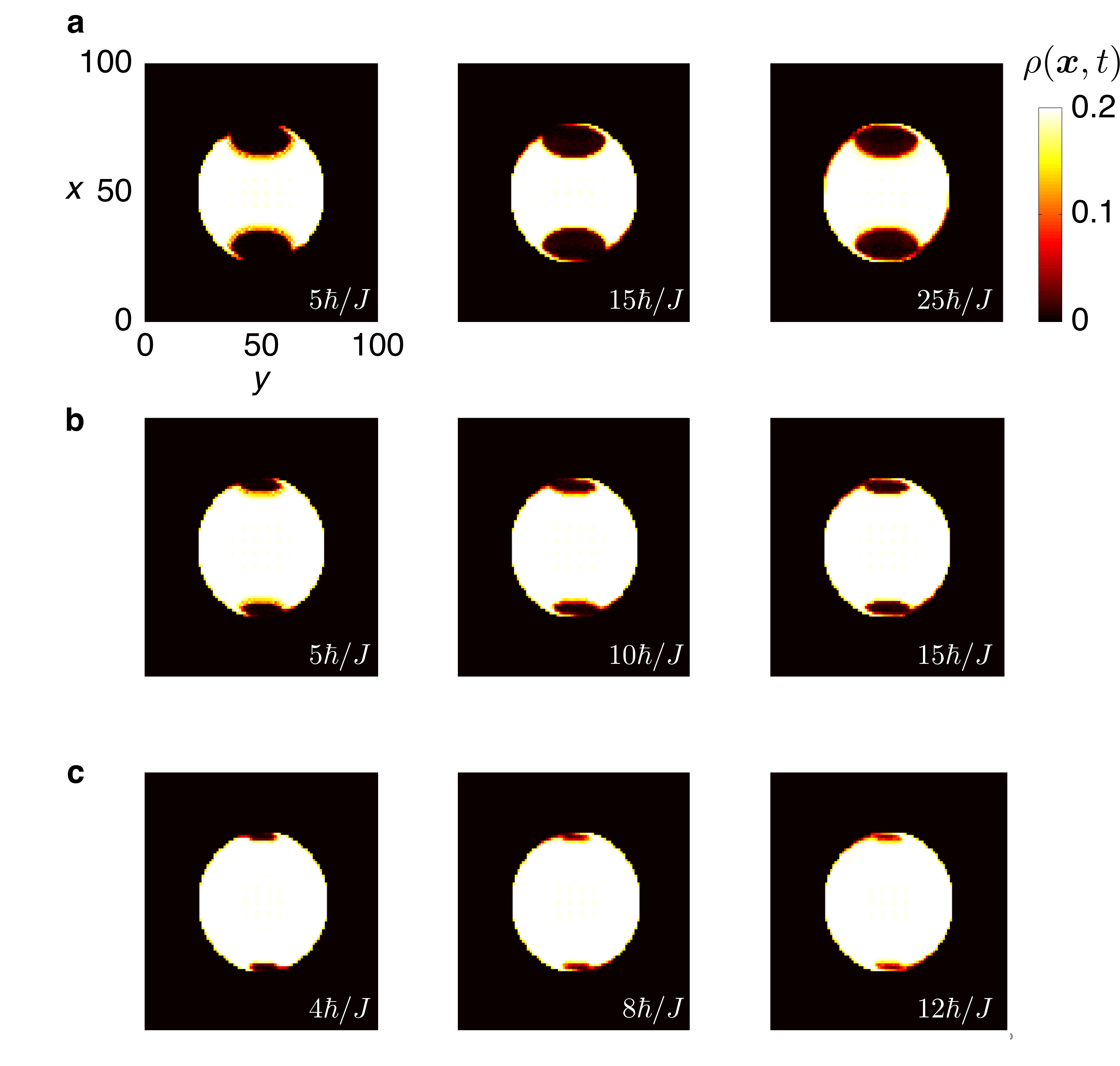}
	\caption{\label{figuresmall}  {\bf Effects of the holes size.} The spatial density $\rho (\bs x , t)$ for $\Phi=1/5$, $E_{\text{F}}=-1.5 J$, $r_0=27a$ and $\gamma=\infty$. The holes are described by Eq. \eqref{ellipses} and correspond to {\bf a,} $b=4$, {\bf b,} $b=8$, and {\bf c,} $b=16$. Note that the chiral motion of the edge states is still visible, even in the limit of tiny perturbative walls $V_{\text{hole}}$.}
\end{figure}

The geometry produced by these potentials is particularly suitable to emphasize the edge states motion at time $t>0$, as they encircle the regions initially surrounded by the elliptical walls. However, in this exotic geometry, the large holes correspond to a massive distortion of the standard circular system. It is thus interesting to study the validity of our method, as we progressively reduce the size of the holes. Here, we  demonstrate that our scheme still shows a clear manifestation of the edge states, even in the limit where the walls $V_{\text{hole}}$ only represent a small perturbation of the system, in the vicinity of its circular boundary. To analyze this, we propose to calculate the time evolution of the density $\rho (\bs x,t)$ after removing the elliptical walls delimited by the more general equations
\begin{equation}
\bigl (x \pm (b-1) \, r_0/b \bigr)^2 + \bigl (y/\sqrt{b} \bigr)^2 =  (r_0/b)^2, \label{ellipses}
\end{equation}
which can be made arbitrarily small ($b\gg 1$), while maintaining the smooth contact with $V_{\text{conf}} (r)$ at the poles. The results are shown in Fig. \ref{figuresmall} for the flat-band configuration previously studied in Fig. \ref{figure4} (for $b=2$), but with smaller initial holes $b=4,8,16$. This picture emphasizes the fact that our scheme still allows to detect the chiral motion of the edge states, in the limit of small perturbative walls $V_{\text{hole}}$. However, we stress that it is crucial to prohibit any broadening of the edge state signal in the perturbative regime $b \gg 2$, which necessarily requires the use of an extremely abrupt external potential $V_{\text{conf}}$ with $\gamma \sim \infty$. Moreover, considering smaller holes also demands to further reduce the bulk dispersion, which can be achieved by considering the quasi-flat-band configuration $\Phi=1/5$, see Fig. \ref{figuresmall}, or by exploiting the ``opposite-flux'' or ``edge-filter'' methods. \\

\begin{figure}[h!]
	\centering
	\includegraphics[width=1.\columnwidth]{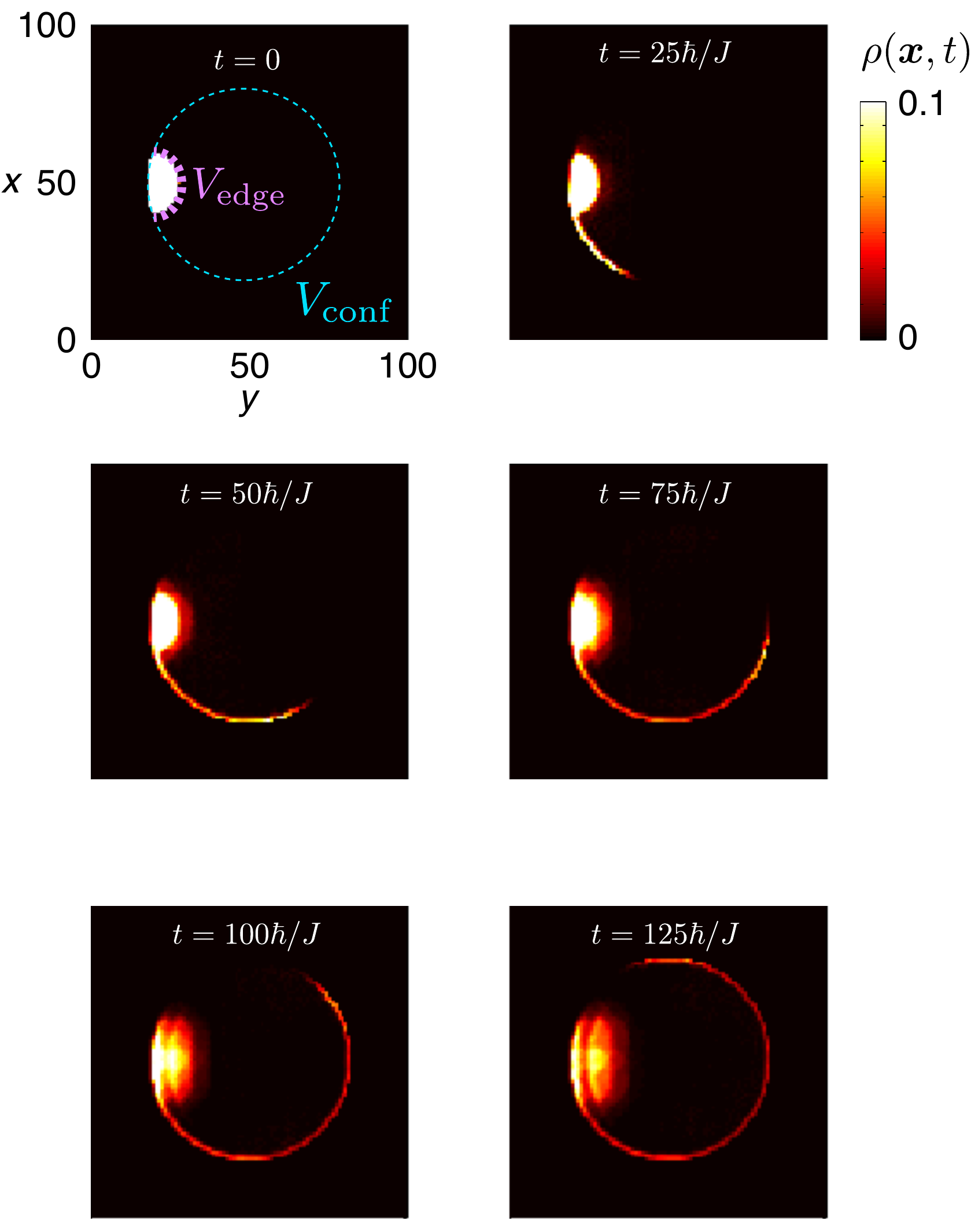}
	\caption{\label{figureedge}  {\bf Preparing the atomic cloud close to the edge.} The spatial density $\rho (\bs x , t)$ for $\Phi=1/5$, $E_{\text{F}}=-1.5 J$, $r_0=32a$ and $\gamma=\infty$. The system is initially confined close to the circular edge at $r=r_0$ by an abrupt potential wall $V_{\text{edge}}$. After releasing the wall $V_{\text{edge}}$, the edge states propagate along the circular edge delimited by $V_{\text{conf}} (r)$, highlighted in the top-left picture by a blue dotted circle. }
\end{figure}

Another relevant configuration is obtained by replacing the constraining potentials $V_{\text{hole}}$ by a spacious wall $V_{\text{edge}}$ that initially confines the entire atomic cloud to a small region located in the vicinity of the circular edge delimited by $V_{\text{conf}} (r)$, see Fig. \ref{figureedge}. After releasing the wall $V_{\text{edge}}$ at time $t=0$, the edge states propagate along the circular edge delimited by $V_{\text{conf}} (r)$, while the bulk states evolve towards the center of the trap. This strategy, which largely improves the edge/bulk ratio, is particularly efficient for dispersionless systems (e.g. $\Phi=1/5$).

\section{Time evolution of a trivial insulating phase}

\begin{figure}[h!]
	\centering
	\includegraphics[width=1\columnwidth]{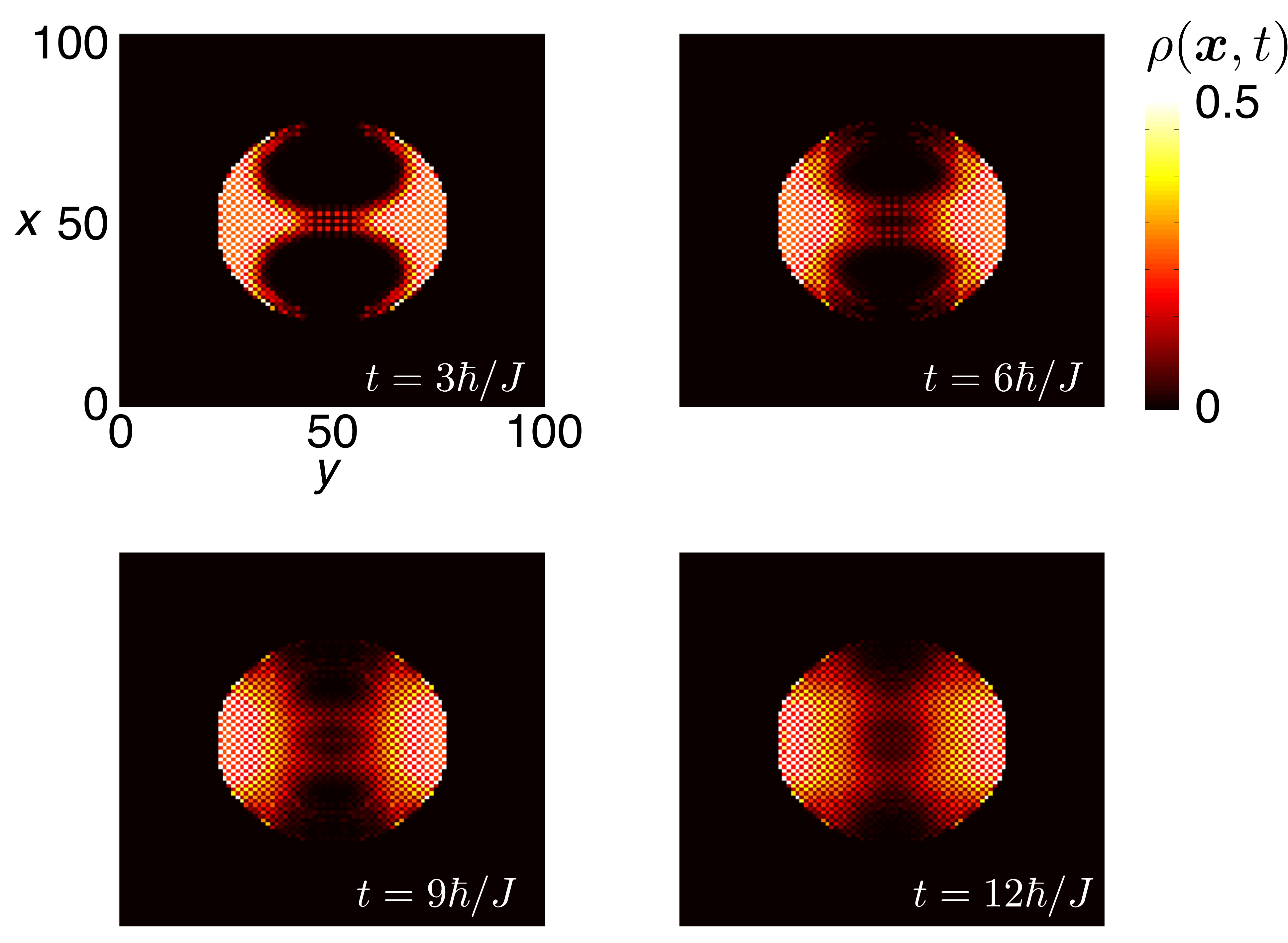}
	\caption{\label{trivialfig}  {\bf Time evolution of a trivial insulating phase.} The spatial density $\rho (\bs x , t)$ for $\Phi=1/2$, $r_0=27a$ and $\gamma=\infty$. The Fermi energy $E_{\text{F}}=0$ is set within a trivial bulk gap ($\nu=0$), opened by a staggered potential of strength $V_{\text{stag}}=J$. Chiral edge states are absent in the time-evolving density, in agreement with the triviality of the Chern number $\nu=0$.}
\end{figure}

The chiral motion of the edge states shown in Figs.~\ref{figure3} and \ref{figure4} is a signature of the non-trivial Chern number $\nu = -1$ (see main text). As illustrated in Fig. \ref{figure5}, reversing the sign of the magnetic flux $\Phi \rightarrow - \Phi$ leads to an opposite chirality, in agreement with the fact that the Chern number also changes its sign under the transformation. Here, we further demonstrate that the edge-states motion, visible in the time-evolving density $\rho (\bs x,t)$, can be unambiguously attributed to the non-triviality of the Chern number. We consider the same system, but in a configuration characterized by a \emph{trivial} Chern number $\nu=0$. This configuration is produced in the following way: (i) We set the flux to the value $\Phi=1/2$, which leads to a gapless bulk energy spectrum displaying two Dirac cones. (ii) We add a staggered potential along both spatial directions, with alternating on-site energies $\pm V_{\text{stag}}$, which opens a bulk gap around $E=0$. This gap is trivial in the sense that the lowest band is associated with a zero Chern number $\nu=0$, and therefore, edge states are unexpected in this configuration. We represent in Fig.~\ref{trivialfig} the time-evolving density, obtained by initially setting the Fermi energy within the trivial gap. This figure \ref{trivialfig}, which is to be compared with Fig.~\ref{figure3}, shows (i) the \emph{non-chiral} dynamics of the bulk states initially occupying the lowest band, and (ii) the absence of chiral edge states. 

\end{document}